\documentclass[aps,twocolumn,prb,amsmath,amssymb]{revtex4-2}

\usepackage{amsmath}
\usepackage{amssymb}
\usepackage[varg]{txfonts}
\usepackage{color}
\usepackage[colorlinks]{hyperref}
\usepackage{tikz}
\usetikzlibrary{shapes}
\usepackage{xcolor}

\usepackage{graphicx}
\usepackage{bm}

\makeatletter
\newcommand\bigDiamond{\mathop{\mathpalette\bigDi@mond\relax}}
\newcommand\bigDi@mond[2]{\vcenter{\hbox{\m@th \scalebox{\ifx#1\displaystyle 2\else1.2\fi}{$#1\Diamond$}}}}
\newcommand{\RNum}[1]{\uppercase\expandafter{\romannumeral #1\relax}}

\def\XXint#1#2#3{{\setbox0=\hbox{$#1{#2#3}{\int}$}
    \vcenter{\hbox{$#2#3$}}\kern-.5\wd0}}

\def\be{\begin{equation}}
\def\ee{\end{equation}}

\def\bi{\begin{itemize}}
    \def\ei{\end{itemize}}
\def\bn{\begin{enumerate}}
    \def\en{\end{enumerate}}
\def\bea{\begin{eqnarray}}
\def\eea{\end{eqnarray}}
\newcommand{\bpm}{\begin{pmatrix}}
    \newcommand{\epm}{\end{pmatrix}}

\def\ba{\begin{array}}
    \def\ea{\end{array}}
\def\bd{\begin{displaymath}}
\def\ed{\end{displaymath}}

\renewcommand{\imath}{\hspace{1pt}\mathrm{i}\hspace{1pt}}

\renewcommand{\vec}{\mathbf}
\renewcommand{\Re}{\mathop{\mathrm{Re}}\nolimits}
\renewcommand{\Im}{\mathop{\mathrm{Im}}\nolimits} 

\begin{document}

\title{Phase transition and fractionalization in superconducting Kondo lattice model}

\author{Fatemeh Mohammadi}
\affiliation{Department of Physics, Sharif University of Technology, Tehran 14588-89694, Iran}

\author{Amirhossein Saedpanah}
\affiliation{Department of Physics, Sharif University of Technology, Tehran 14588-89694, Iran}

\author{Abolhassan Vaezi}
\affiliation{Department of Physics, Sharif University of Technology, Tehran 14588-89694, Iran}

\author{Mehdi Kargarian}
\email{kargarian@sharif.edu}
\affiliation{Department of Physics, Sharif University of Technology, Tehran 14588-89694, Iran}

\begin{abstract}
Topology, symmetry, electron correlations, and the interplay between them have formed the cornerstone of our understanding of quantum materials in recent years and are used to identify new emerging phases. While the first two give a fair understanding of noninteracting and, in many cases, weakly interacting wave function of electron systems, the inclusion of strong correlations could change the picture substantially. The Kondo lattice model is a paradigmatic example of the interplay of electron correlations and conduction electrons of a metallic system, describing heavy fermion materials and also fractionalized Fermi liquid pertaining to an underlying gauge symmetry and topological orders. In this work, we study a superconducting Kondo lattice model, a network of 1D Kitaev superconductors Kondo coupled to a lattice of magnetic moments. Using slave-particle representation of spins and exact numerical calculations, we obtain the phase diagram of the model in terms of Kondo coupling $J_K$ and identify a topological order phase for $J_{K}<J_{K}^c$ and a Kondo compensated phase for $J_{K}>J_{K}^c$, where $J_{K}^c$ is the critical point. Setting the energy scales of electron hopping and pairing to unity, the mean-field theory calculations achives $J_{K}^c=2$ and in exact numerics we found $J_{K}^c\simeq 1.76$, both of which show that the topological order is a robust phase. We argue that in terms of slave particles, the compensated phase corresponds to an invertible phase, and a Mott insulating transition leads to a topological order phase. Further, we show that in the regime $J_{K}<J_{K}^c$ in addition to the low-energy topological states, a branch of subgap states appears inside the superconducting gap.
\end{abstract}

\maketitle

\section{Introduction}
The Kondo lattice model, a lattice of magnetic ions interacting with a host of conduction electrons gas, is a prototype example of strongly interacting systems with localized $d$- or $f$- electron orbitals, exhibiting a plethora of correlated phases and quantum criticality \cite{Continentino1994, sachdev:Book2011, Hertz1976, Millis1993, Qimiao2001, Coleman2002, Coleman:Nature2005}. Due to strong Coulomb interaction between electrons in latter orbitals, the charge fluctuations are suppressed, and hence the localized state is described by a quantum spin $\mathbf{S}$. The spin interacts antiferromagnetically with the spin of electrons through the exchange coupling $J_K$ known as Kondo coupling, which sets the Kondo temperature $T_{K}\propto e^{-1/\rho(0)J_{K}}$, with $\rho(0)$ as the density of states at the Fermi surface. The magnetic moments can also interact with each other either through the Fermi surface of electrons, a.k.a., the RKKY interaction \cite{Ruderman1954, Kasuya1956, Yosida1957}, or directly via Heisenberg spin exchange interaction. The latter is known as the Heisenberg-Kondo lattice model \cite{Coleman1989}. Notwithstanding strong correlations on magnetic sites, the Kondo lattice model describes a heavy Fermi liquid material \cite{Hewson1993, Coleman:Book2007, Coleman:Book2015} with the significance of having a large Fermi surface encompassing both conduction electrons and local magnetic moments satisfying the Luttinger theorem \cite{Luttinger1960}.        

There is yet a more exotic phase where the magnetic moments do not form ordered states and are effectively mean-field decoupled from the conduction band, resizing the Fermi volume to include only the conduction electrons. For these nonmagnetic metallic states, known as fractional Fermi liquid, the electron states near the Fermi surface are characterized by sharp quasiparticles carrying spin $S=1/2$ and electron charge $e$. Besides, there exists a branch of spin $S=1/2$ neutral excitations associated with the deconfined phase of the underlying gauge theory describing the spin liquid state of magnetic moments. The simplest gauge group has a discrete $\mathbb{Z}_2$ symmetry possessing gapped topological vison excitations. The bottom line here is that the topological order in the Heisenberg-Kondo lattice model could lead to fractionalization, i.e., the system possesses both electrically charged and neutral excitations \cite{Senthil:PRL2003}. Also, the topological order and fractionalization appear as a natural description of strongly correlated systems with superconducting pairing fluctuations \cite{ Senthil:prb1999}. 


The spectrum of magnetic moments in conventional superconductors has been widely studied. A magnetic moment in a superconductor leads to some subgap states in the superconducting gap known as Yu-Shiba-Rusinov (YSR) \cite{Yu1965, Shiba1968, Rusinov1968} states. Naively the superconducting gap suppresses the metallic Kondo screening cloud with a typical size of screening length $\xi_{K}\sim v_{F}/T_K$, with $v_F$ as Fermi velocity, due to the absence of particle-hole excitations within the superconducting gap. The formation of YSR-bound states, however, implies that the magnetic moment is partially screened, and the spin of the magnetic moment forms singlets with Bogoliubov quasiparticles.

The study of an analogue of Kondo lattice model in a superconducting medium is the main goal of this paper. Unlike the local magnetic impurities, which have been studied extensively, up to our best knowledge, the superconducting Kondo lattice model is limited to a few studies, which we briefly review below. A chain of magnetic ions of iron in proximity to an s-wave superconductor has been experimented in Ref.[\onlinecite{Yazdani2014}] to realize the 1D Kitaev superconductor \cite{Kitaev2001}. The main result is the observation of zero-bias peaks in tunneling transport at the ends of the chain with a possible explanation in terms of Majorana bound state. Although other alternative explanations based on the mid-gap bound states were given shortly after the experimental observation, a set of recent works argue that a topological phase transition may occur in a band of YSR states. The topological phase is described by an effective $p$-wave pairing and hence may account for the zero states at the ends of the chain \cite{Pientka2013}.

Another example of superconducting Kondo lattice model is a network of 1D Kitaev superconducting chains Kondo coupled to a lattice of magnetic ions \cite{Hsieh2017}, which is the main focus of our work in this paper. The model provides a fertile ground for merging topology and strong correlations where topologically ordered phases arise. In the limit of weak Kondo coupling $J_{K}\ll \Delta$, where $\Delta$ is the superconducting gap, a simple degenerate perturbation theory generates a spin model whose low-energy states are described by $\mathbb{Z}_2$ topological order. This construction provide a framework to realize more exotic topological orders. For instance, by designing magnetic ions on a honeycomb lattice, we recently introduced topological models with local $\mathbb{Z}_2\times \mathbb{Z}_2$ gauge symmetry \cite{Mohammadi2022}, and a lattice of Majorana Cooper boxes inducing $\mathbb{Z}_2$ topological order. In all of these models, it seems a combination of nontrivial band topology of superconductor and strong correlations of Kondo couplings lead to more exotic phases with topological orders at least perturbatively in $J_K$.

In this paper, we ask the following questions. Does the topological phase arise only at small values of Kondo coupling $J_K$? Is there a phase transition out of the topologically ordered phase at some critical coupling $J^{c}_{K}$? The main motivation for asking these questions is that, while the perturbation theory is valid at small $J_{K}$ with emerging topological order, at the limit of large Kondo coupling, one expects that the formation of Kondo singlets between the spin of superconducting electrons and magnetic moments becomes the groundstate of the system which is a trivial phase. Therefore, there should be at least one phase transition by increasing $J_{K}$. We use slave-particle representation of spin of local moments to derive a fully fermionic model being well suited for large-N treatment and mean-field theory. Also, we use exact numerical calculations using the density matrix renormalization group (DMRG) on finite-size lattices to explore the phase transition further.

Let us briefly review the main findings of this work: (i) we diagnose a quantum phase transition occurring at a rather large value of Kondo coupling $J^{c}_{K}$, and hence, (ii) we conclude that the $\mathbb{Z}_2$ topological order phase is rather robust beyond the perturbation limit, (iii) we recognize the trivial phase at $J_K>J^{c}_{K}$ as an invertible phase \cite{Hsieh2016}, and (iv) we argue that the phase transition can be understood as a Mott insulating transition from the invertible phase to a topological order, (v) a branch of YSR bound states appears within the superconducting gap for $J_K<J^{c}_{K}$, whose energy vanishes at the critical point, (vi) the exact numerical calculation can faithfully capture the phase transition and the topological order.

The paper is organized as follows. In Sec.~\ref{model} we introduce the superconducting Kondo lattice model and describe the model by an effective mean-field Hamiltonian. We analyze the details of the phase diagram of the model in Sec.~\ref{phase_diagram}. A quasi-one-dimensional model with $\mathbb{Z}_2$ topological order is proposed in Sec.~\ref{quasi_one_dimension} to examine the possible phases using the DMRG numerical calculations. Sec.~\ref{conclusions} summarizes and concludes.

\section{Superconducting Kondo Lattice Model}\label{model}
\begin{figure}[t]
\center
\includegraphics[width=\linewidth]{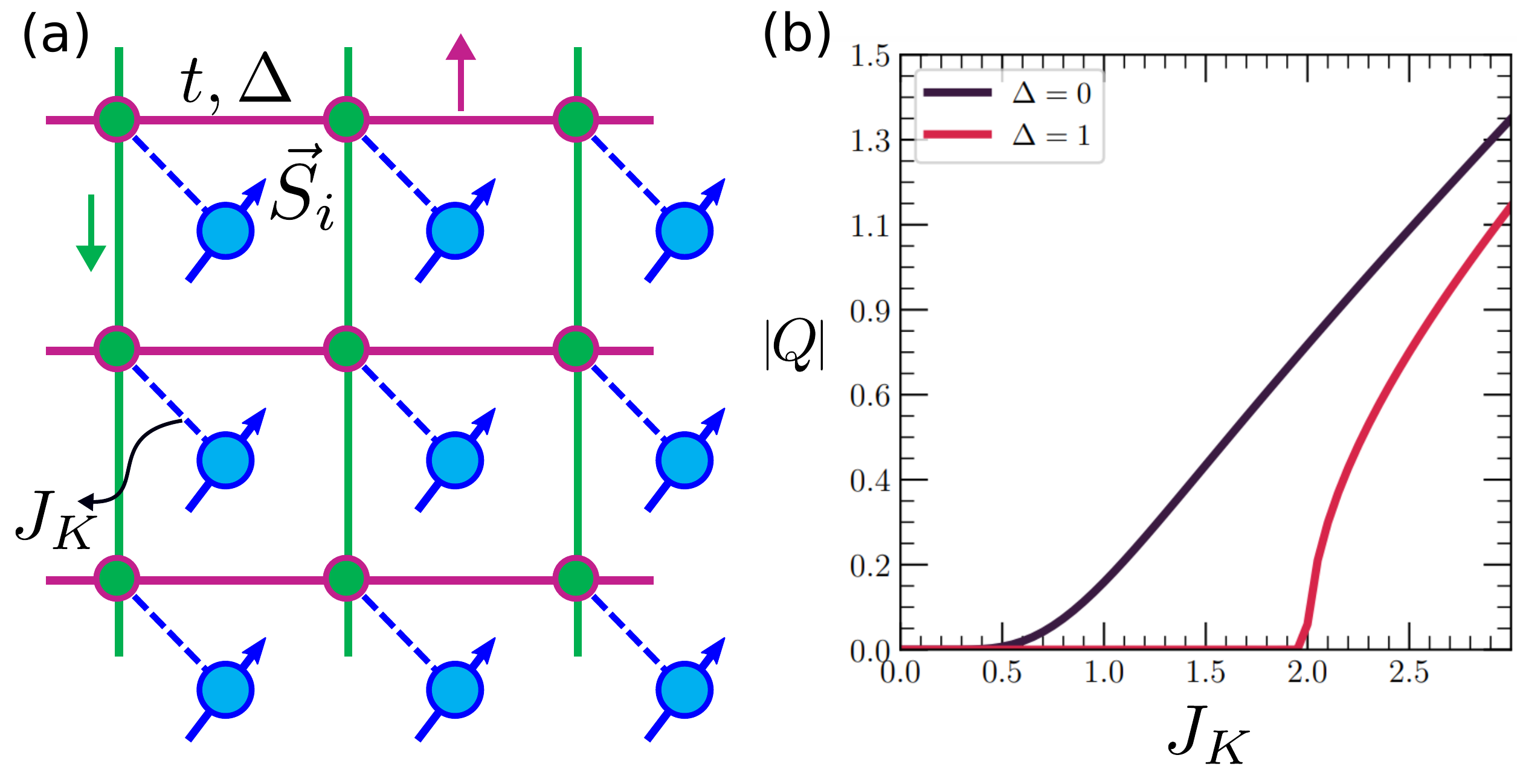}
\caption{(a) The superconducting Kondo lattice model. Red (green) wires are superconducting Kitaev chains with spin-up (down) electrons. The superconductor is coupled antiferromagnetically to magnetic moments shown as dashed lines to blue circles. (b) The mean-field parameter $|Q|$ as a function of Kondo exchange coupling $J_K$. The black line corresponds to $\Delta = 0$, corresponding to a metallic Kondo lattice model with marginally relevant Kondo coupling. The red line indicates the variation of $|Q|$ for the superconducting Kondo lattice model with $\Delta = 1$. The system undergoes a quantum phase transition at $J_K^c=2$. Two phases are uncovered: (1) an uncondensed phase with $|Q|=0$, which is topologically ordered at low energies with underlying $\mathbb{Z}_2$ gauge symmetry, (2) a condensed phase with $|Q|\neq0$ featuring an invertible topological phase.\label{topological_superconductor_lattice}}
\end{figure}

Our starting point is the model adopted in Ref.[\onlinecite{Hsieh2017}], a superconductor coupled to magnetic moments. The superconducting model is a lattice composed of one-dimensional spin-polarized $p$-wave Kitaev chains so that the horizontal chains carry spin-up electrons and vertical chains carry spin-down electrons. Two species of electrons meet at lattice sites, as shown in Fig.~\ref{topological_superconductor_lattice}. The superconducting Hamiltonian reads as

\begin{equation}\label{Kitaev_chain}
\begin{split}
H_{\mathrm{sc}}[c^{\dagger},c] =\sum_{j} \left(-t c_{j, \alpha}^{\dagger}c_{j + \hat{a}_{\alpha}, \alpha} + \Delta c_{j, \alpha} c_{j + \hat{a}_{\alpha}, \alpha} + \mathrm{h.c.} - \mu  c_{j, \alpha}^{\dagger}c_{j , \alpha} \right),
\end{split}
\end{equation}
where $\alpha = \uparrow, \downarrow$ are spin's degrees of freedom, $t$, and $\Delta$ are hopping amplitude and superconducting pairing potential, respectively, and $\mu$ is the chemical potential.The sum runs over the sites $j\equiv \mathbf{R}_j$ of a two-dimentional lattice with $\hat{a}_{\uparrow}=\hat{x}$ and $\hat{a}_{\downarrow}=\hat{y}$, and the summation convention over repeated spin indices $\alpha$ is assumed implicitly throughout. Each chain is in a topological superconducting phase when $\mu<|2t|$ \cite{Kitaev2001}. The superconducting chains are antiferromagnetically coupled to a lattice of otherwise free magnetic moments. The superconducting Kondo lattice model (SKLM) then reads as

\begin{equation}\label{H_SKL}
H_{\mathrm{SKLM}} =H_{\mathrm{sc}}[c^{\dagger},c] + \frac{J_K}{2} \sum_{j} \vec{S}_j \cdot (c_{j, \alpha} ^ {\dagger} \vec{\sigma}_{\alpha, \beta} c_{j,\beta}).
\end{equation}

To study the phase diagram of this model, we use the slave-fermion representation of spin operators,

\begin{equation}\label{parton_construction}
\vec{S}_j = \frac{1}{2} f_{j,\alpha}^{\dagger}\tau_{\alpha, \beta} f_{j,\beta}, \quad  f_{j,\alpha}^{\dagger} f_{j,\alpha} = \frac{N_f}{2}
\end{equation}
where $f$'s are neutral fermionic operators, the spinons, $\left\{f_{\alpha},f^{\dagger}_{\beta}\right\}=\delta_{\alpha\beta}$ and the summation convention on repeated indices is implicit. Keeping an eye into the use of saddle-point approximation in next sections, in writing \eqref{parton_construction} we promoted the original SU(2) representation of spin to include large number of flavors $N_{f}$, i.e., $\alpha, \beta = 1, 2, \cdots, N_f$. The second expression in \eqref{parton_construction} is a constraint on the number of flavors on each site (e.g., $N_f=2$ for $S=1/2$ spins), the number of spinons needed to represent the spin. As usual the constraint is imposed by a Lagrange multiplier in the path integral formulation as $i \sum_{j}\lambda_j \left( f_{j, \alpha}^{\dagger} f_{j, \alpha} - \frac{N_f}{2}\right)$. We proceed by writing the Kondo coupling in \eqref{H_SKL} as
\begin{equation}\label{Kondo_int}
-\frac{J_{K}}{N_f}\sum_{j}c^{\dagger}_{j,\alpha}f_{j,\alpha}f^{\dagger}_{j,\beta}c_{j,\beta},
\end{equation} 
where we ignored an unimportant single-particle term which is absorbed into the chemical potential of the superconductor. Using Hubbard-Stratonovich fields $Q$, the interaction can be decomposed, and the total action becomes,
\begin{widetext}
\begin{equation}\label{action}
S=\int d\tau\left\{c^{\dagger}\partial_{\tau}c+ f^{\dagger}\partial_{\tau}f +H_{sc}[c^{\dagger},c]+ H_{cf}[f, c, Q]-i \sum_{j} \lambda_j\left( f_{j, \alpha}^{\dagger} f_{j, \alpha} - \frac{N_f}{2}\right)  \right\}, 
\end{equation}
\end{widetext}    
where
\begin{equation}\label{Hcf}
H_{cf}[f, c, Q]=\sum_{j}\left(\frac{N_f}{J_{K}}|Q_j|^2-Q_jf^{\dagger}_{j,\beta}c_{j,\beta}-Q^{*}_jc^{\dagger}_{j,\beta}f_{j,\beta} \right).
\end{equation}

In the limit of large $N_f$, the model is well suited for mean-field approximation and seeking saddle-point solutions. Note that we could also decompose the interaction term in pairing channel by introducing fields $P\equiv f^{\dagger}_{i\alpha}c^{\dagger}_{i\beta}$. However, we found that the pairing fields vanishes for all $J_K$ in the mean-field approximation, hence ignoring it in \eqref{action}. Assuming a uniform ansatz at the saddle point, $Q_{j}\rightarrow Q$ and $\lambda_j\rightarrow i\lambda$, The mean-field Hamiltonian is given by

\begin{equation}\label{H_MF}
H_{\mathrm{MF}}=H_{\mathrm{sc}}[c^{\dagger},c]+H_{Q}[f,c]+\lambda \sum_{\vec{k},\alpha} \left( f_{\vec{k}, \alpha}^{\dagger} f_{\vec{k}, \alpha}  -\frac{N_f}{2}\right),
\end{equation} 
 where
\begin{equation}\label{Hsc}
H_{\mathrm{sc}}[c^{\dagger},c]=\sum_{\vec{k}, \alpha}\left(\varepsilon(k_{\alpha})  c^{\dagger}_{\vec{k},\alpha}c_{\vec{k},\alpha}+\frac{\Delta(k_{\alpha})}{2}c_{-\vec{k},\alpha}c_{\vec{k},\alpha}+\mathrm{h.c.}\right)
\end{equation}
with $\Delta(k_{\alpha})=2i\Delta\sin(k_{\alpha})$ an dispersion $\varepsilon(k_{\alpha})=-2t\cos(k_{\alpha})-\mu$ for spin-up (spin-down) electrons moving along $k_{\uparrow}\equiv k_x$ ($k_{\downarrow}\equiv k_y$) directions, and  
\begin{equation}\label{Hfc}
H_{Q}[f,c]=\sum_{\vec{k},\alpha}\left(\frac{N_f}{J_K}|Q|^2-Qf^{\dagger}_{\vec{k},\alpha}c_{\vec{k},\alpha}-Q^{*}c^{\dagger}_{\vec{k},\alpha}f_{\vec{k},\alpha} \right).
\end{equation}

The saddle-point equations determining $Q$ and $\lambda$ read as
\begin{equation}\label{saddle_point_equation}
\begin{split}
&Q = \frac{J_K}{N_fN}\sum_{\vec{k}}\langle c_{\vec{k}, \alpha}^{\dagger}f_{\vec{k}, \alpha} \rangle \\
&\frac{1}{2} = \frac{1}{N_fN}\sum_{\vec{k}}\langle f_{\vec{k}, \alpha}^{\dagger}f_{\vec{k}, \alpha} \rangle,
\end{split}
\end{equation}
where $N$ is the number of lattice sites. Our remaining task is to solve the saddle-point equations \eqref{saddle_point_equation} self-consistently using $H_{\mathrm{MF}}$ to obtain the phase diagram of the model. 

Before closing this section, let us review a metallic version of the model \eqref{H_SKL} without pairing, namely, we set $\Delta=0$. This is the usual Kondo lattice model of conduction band coupled to localized spins. The mean-field Hamiltonian is the same as in \eqref{H_MF} except that the superconducting hamiltonian is replaced by noninteracting electrons $H_{0}=\sum_{\mathbf{k}} \varepsilon(k_{\alpha}) c^{\dagger}_{k_{\alpha}}c_{k_{\alpha}}$.
 

It is convenient to write the mean-field equations in \eqref{saddle_point_equation} in terms of Green's functions $\mathcal{G}_{cf, \sigma}(\vec{k}, \tau=0^{-}) = - \lim_{\tau \rightarrow 0^-} \langle T_{\tau} f_{\vec{k}, \alpha}(\tau) c_{\vec{k}, \alpha}^{\dagger}(0)\rangle$ and $\mathcal{G}_{f, \sigma}(\vec{k}, \tau=0^{-}) = - \lim_{\tau \rightarrow 0^-} \langle T_{\tau} f_{\vec{k}, \alpha}(\tau) f_{\vec{k}, \alpha}^{\dagger}(0)\rangle$:

\begin{equation}\label{self_consistent_eq}
\begin{split}
&\frac{Q}{J_K} = \frac{T}{2N} \sum_{\vec{k}, ik_n}\mathcal{G}_{cf, \sigma}(\vec{k}, ik_n)e^{ik_n0^{-}} \\
&1 = \frac{T}{N} \sum_{\vec{k}, ik_n} \mathcal{G}_{f, \sigma}(\vec{k}, ik_n)e^{ik_n0^{-}}.
\end{split}
\end{equation}	

Here $k_n = (2n + 1)\pi T$ ($T$ is the temperature) is the fermionic Matsubara frequencies and  

\begin{equation}\label{g_f_final}
\mathcal{G}_{f, \sigma}(\vec{k}, ik_n) =\left[ ik_n - \lambda - \frac{|Q|^2}{ik_n - \varepsilon(k_{\sigma})}\right]^{-1},
\end{equation}

\begin{equation}\label{g_fc_final}
\begin{split}
\mathcal{G}_{fc, \sigma}(\vec{k}, ik_n) = &\frac{-Q}{ik_n - \lambda} \left[ik_n - \varepsilon(k_{\sigma}) - \frac{|Q|^2}{ik_n - \lambda} \right]^{-1}.
\end{split}
\end{equation}

The saddle-point equation for $Q$ admits a nonzero solution. The results is shown in Fig.~\ref{topological_superconductor_lattice}(b). It is seen that the order parameter behaves like much studied the Kondo lattice models adopted in studies of heavy fermion materials with an immediate consequence that the Kondo coupling $J_K$ is marginally relevant, the $Q$ exponentially vanishes as $J_{K}\rightarrow0$.


One salient feature of the metallic Kondo lattice model is that the Fermi surface now includes both metallic electrons and spins. Indeed the wave functions of localized spins can leak inside the Fermi sea, implying that the spins now become part of the Fermi liquid. In terms of spinon representation of spins in \eqref{parton_construction}, the latter argument suggests that the spinons acquire charge upon condensation of $Q$ so that in the computation of the Fermi volume, both types of fermions, $c$ and $f$, should be counted. In a work by Coleman \emph{et al.} in Ref.[\onlinecite{Coleman:prb2005}], the authors describe this phenomenon as the formation of charged heavy electrons. In the next section, we use this latter notion to describe the phase transition to a Mott insulating phase by stripping off the charge. One remark on gauge symmetry is in order. The spinon representation in \eqref{parton_construction} is clearly redundant: under the U(1) gauge transformation $f_{j,\alpha} \rightarrow f_{j,\alpha} e^{i\phi_{j}(\tau)}$, the spin remains invariant, implying that the low-energy theory has a gauge structure with respect to this gauge transformation. In particular, we notice that the field $Q$ carries the same charge under such U(1) gauge transformation. However, when $Q$ is condensed and acquires a nonzero value, the U(1) gauge fluctuations become massive due to Higgs mechanism (see the Appendix \ref{Higgs} for details) and thus the mean-field solution remains stable.

\section{Phase diagram of Superconducting Kondo Lattice Model}\label{phase_diagram}
We now proceed by solving the mean-field equations in \eqref{saddle_point_equation} for the superconducting model described in the preceding section. The equations can be written in terms of new Green's functions, including anomalous terms associated with superconductivity. The expressions are lengthy and are given in Appendix \ref{Green_SC}. Alternatively one can directly solve \eqref{saddle_point_equation} self-consistently using the eigenstates of \eqref{H_MF}. For numerical solution we set $t=\Delta=1$ and $\mu=0$. The zero chemical potential implies the particle-hole symmetry of the nonsuperconducting model at half-filling, and then $\lambda=0$ solves the second equation in \eqref{saddle_point_equation}. The values of $Q$ versus Kondo coupling $J_K$ are shown in Fig.~\ref{topological_superconductor_lattice}(b). Unlike the metallic case, as discussed at the end of the preceding section, there is a phase transition in the superconducting Kondo lattice model with a critical point at $J_K^{c}$ ($=2$ for the aforementioned parameters). The order parameter $Q$ develops nonzero values beyond the critical Kondo coupling stressing that the $J_K$ is no longer marginal.

To derive the critical point for generic values of $t$ and $\Delta$, we can evaluate the fluctuations of the field $\hat{Q}$. The Kondo interaction in \eqref{Kondo_int} can be written as 

\begin{equation}
H_{K}=-\frac{J_K}{N_f}\sum_{\mathbf{q}}\hat{Q}(\mathbf{q})\hat{Q}^{\dagger}(\mathbf{q}),~~~\hat{Q}(\mathbf{q})=\sum_{\mathbf{k},\alpha}c^{\dagger}_{\mathbf{k}-\mathbf{q},\alpha}f_{\mathbf{k},\alpha}. 
\end{equation} 

We have to evaluate the $Q$-susceptibility in the imaginary-time formulation
\begin{equation}
\chi_{Q}(\mathbf{q},\tau)=-\left\langle T_{\tau}\hat{Q}(\mathbf{q},\tau)\hat{Q}^{\dagger}(\mathbf{q},0)\right\rangle
\end{equation}
and the average is taken over interacting many-body states. We are aiming at the evaluation of $Q$-susceptibility in the RPA approximation in the Matsubara domain

\begin{equation}\label{chi_RPA}
\chi^{\mathrm{RPA}}_{Q}(\mathbf{q},iq_n)=\frac{\chi^0_{Q}(\mathbf{q},iq_n)}{1-V_0\chi_{Q}^0(\mathbf{q},iq_n)},
\end{equation}
where $V_0=-J_K/N_f$, $q_n=2n\pi T$ is the bosonic Matsubara frequency. Here, the bare $Q$-susceptibility $\chi^0_{Q}(\mathbf{q},iq_n)$ reads as,

\begin{equation}\label{chi0}
\chi^0_{Q}(\mathbf{q},iq_n)=T\sum_{\mathbf{k},\alpha}\sum_{m}\mathcal{G}^{0}_{c,\alpha}(\mathbf{k},ik_m)\mathcal{G}^{0}_{f,\alpha}(\mathbf{k}+\mathbf{q},ik_m+iq_n),
\end{equation}
with $\mathcal{G}^{0}_{c,\alpha}$ and $\mathcal{G}^{0}_{f,\alpha}$ as bare Green's functions of superconductor and spinons, respectively, reading as,

\begin{equation}
\mathcal{G}^{0}_{c,\alpha}(\mathbf{k},ik_m)=\left(ik_m-\varepsilon(k_{\alpha})-\frac{|\Delta(k_{\alpha})|^2}{ik_m+\varepsilon(k_{\alpha})} \right)^{-1},
\end{equation} 
\begin{equation}
\mathcal{G}^{0}_{f,\alpha}(\mathbf{k},ik_m)=\frac{1}{ik_m-\lambda}.
\end{equation} 

Since the spinons are dispersionless, the $\chi^0_{Q}(iq_n)\equiv\chi^0_{Q}(\mathbf{q},iq_n)$ and hence $\chi^{\mathrm{RPA}}_{Q}(iq_n)\equiv\chi^{\mathrm{RPA}}_{Q}(\mathbf{q},iq_n)$ become momentum independent. That is, the correlations are purely local in space and extend only along the time direction. Performing the Matsubara sum in \eqref{chi0}, we obtain

\begin{equation}\label{chi0n}
\chi^0_{Q}(iq_n)=\\ 2\sum_{\mathbf{k},\alpha}\left[\frac{1}{2E(k_{\alpha})}\frac{E(k_{\alpha})-\varepsilon(k_{\alpha})}{iq_n-E(k_{\alpha})}+\frac{1}{2}\frac{iq_n-\varepsilon(k_{\alpha})}{q_n^2+E^2(k_{\alpha})} \right],
\end{equation}
where $E(k_{\alpha})=\sqrt{\varepsilon(k_{\alpha})^2+|\Delta(k_{\alpha})|^2}$ and we set $\mu=\lambda=0$. The prefactor, 2, results from two sets of bubble diagrams with the same value. We can now analytically continue the response function to real frequencies $iq_n\rightarrow \Omega+i0^+$ . At $\Omega=0$, equation \eqref{chi0n} yields  

\begin{equation}\label{chi0_F}
\chi^0_{Q}(0)=-\frac{1}{\pi t}F\left(\pi,\sqrt{1-x^2}\right),
\end{equation}
where $x=\Delta/t$ and $F(\phi,y)=\int_{0}^{\phi}d\theta/\sqrt{1-y^2\sin^2\theta}$ is the incomplete elliptical integral of the first kind. The corresponding RPA response function in \eqref{chi_RPA} can be evaluated accordingly, from which we can deduce the critical coupling $J_K^c$ where the response diverges. It yields,

\begin{equation}\label{Jcritical}
J_K^c=\frac{2\pi t}{F\left(\pi, \sqrt{1-x^2} \right)}.
\end{equation}

For $x=1$, we have $F(\pi, 0)=\pi$ and hence $J_K^c=2$, as mentioned before and shown in Fig.~\ref{topological_superconductor_lattice}(b). For the metallic case, where $x=0$, the function $F$ is logarithmically diverging, which leads to $J_K^c\rightarrow 0$.

Thus, we identify two phases for the superconducting Kondo lattice model $x\neq0$: (i) the condensed phase for $J_K>J_K^c$ with $Q\neq0$ and (ii) the uncondensed for $J_K<J_K^c$ phase $Q=0$. While both phases are topological in some sense, their topological contents are drastically distinct from each other. In the following, we present arguments demystifying the nature of these phases and the phase transition between them.

\subsection{Condensed phase: a topological invertible phase}\label{condensed}
As explained above, the order parameter $Q$ develops nonzero values for $J_K>J_K^c$. $Q\neq0$ allows for tunneling between $c$ and $f$ states, and therefore the condensed phase can be described as $H_{\mathrm{con}}=H_{sc}[c^{\dagger},c]+H_{Q}[f,c]$ with the corresponding Hamiltonians given in \eqref{Hsc} and \eqref{Hfc}, respectively. One can think of this system as a model of two layers, the top layer is the superconducting model described by $H_{sc}[c^{\dagger},c]$ and the bottom layer contains only $f$ fermions with flat dispersion. Condensation of $Q$ simply means that the electron states of the superconducting layer are locally hybridized with the charged $f$ fermions living in the bottom layer. In the absence of hybridization, $Q=0$, the ground state manifold of the system is highly degenerate $|\psi_{sc}\rangle\otimes|\psi_{f}\rangle$, where $|\psi_{sc}\rangle$ is the BSC ground state of superconductor and $|\psi_{f}\rangle$ is the state of $f$ fermions with single site occupation. One can either integrate out the superconductor in path integral formulation or use the degenerate perturbation theory to obtain an effective Hamiltonian describing $f$ fermions. In the limit of $Q\ll \Delta$, we obtain the following Hamiltonian,

\begin{equation}\label{Hf}
H_{f}= -t_f \sum_{j} \left(-f_{j,\alpha}^{\dagger} f_{j + \hat{a}_{\alpha}, \alpha} + f_{j, \alpha} f_{j + \hat{a}_{\alpha}, \alpha} + \mathrm{h.c.}\right),
\end{equation}  
where $t_f=Q^2/4$ and for simplicity we assumed $t=\Delta=1$ in the superconducting Hamiltonian. The top superconducting layer induces superconducting potential between $f$ fermions in the bottom layer. Not only that, it is also a topological superconductor but with the opposite winding number for each chain, i.e., the winding number $W_f=-1$ for $f$ superconductor as opposed to $W_c=+1$ for $c$ superconductor. This is the generic feature of symmetry-protected topological (SPT) phases: an SPT phase is an \emph{invertible} phase, i.e., it induces a ground state with opposite topological invariant in an otherwise trivial free system via the bulk topological proximity effect \cite{Hsieh2016}.

Having identified the $f$ fermions as a topological superconductor, we can now think of the total system $c+f$ as a trivial superconductor with $W=W_c+W_f=0$ invariant. To further justify the triviality of the condensed phase of a typical chain, let us examine the topological properties of the mean-field Hamiltonian $H_{MF}=\frac{1}{2}\sum_{k}\psi^{\dagger}_{k}H(k)\psi_{k}$. The Bloch Hamiltonian is 

\begin{equation}
H(k) = 
 \begin{pmatrix}
  \varepsilon(k) & \Delta(k)^{*} & Q & 0 \\
  \Delta(k) & -\varepsilon(k) & 0 & -Q^* \\
  Q^*  & 0  & \lambda & 0  \\
  0 & -Q & 0 & -\lambda
 \end{pmatrix}
\end{equation}
in the basis $\psi_{k}^{\dagger}=(c^{\dagger}_k,c_{-k},f^{\dagger}_k,f_{-k})$. Now we can evaluate the winding number of occupied states, which is given by sum over Zak phases:

\begin{align}
\gamma=-i\sum_{n\in\mathrm{occ.} }\oint\langle u_{n}(k)|\partial_{k}u_{n}(k) \rangle~dk,
\end{align}    
where $|u_{n}(k) \rangle$ is the eigenstate of Hamiltonian $H(k)|u_{n}(k) \rangle=\varepsilon_n(k)|u_{n}(k) \rangle$. The winding number is then given by $W=\gamma/\pi$. We evaluated the winding number numerically and found that for any $Q\neq0$ it vanishes, $W=0$, justifying our physical picture that the entire system is trivial. Thinking in terms of spins in the limit of large Kondo coupling, the spin of superconducting electrons on each site forms a singlet state with the magnetic moment at the same site, $(|\uparrow_c\Downarrow\rangle - |\downarrow_c\Uparrow\rangle)/\sqrt{2}$. Thus, it is natural to expect that the strongly coupled phase is a trivial many-body singlet phase, the so-called Kondo screened phase arising when the Kondo screening length $\xi_{K}$ is smaller than the superconducting coherence length $\xi_{\mathrm{sc}}$, where $\xi_{\mathrm{sc}}\sim v_{F}/\Delta$. We come back to this point when we discussed the phase in $J_K<J_K^c$ in section Sec.\ref{YSR}.

Before leaving this subsection, we would like to comment on the superconducting Hamiltonian in \eqref{Hf} asserting that its general form could be inferred from the symmetry argument. The $c$-superconductor breaks the $U(1)$ symmetry, i.e., the Hamiltonian in \eqref{Kitaev_chain} is $\mathbb{Z}_2$ symmetric under transformation $c_i\rightarrow \epsilon_i c_i$ with $\epsilon_i=\pm1$. On the other hand, the condensation of $Q$ breaks the $U(1)$ gauge symmetry as we discussed at the end of Sec.\ref{model}. This means that for the hybridization Hamiltonian to be invariant, the $f$ fermions should be transformed in the same manner, $f_i\rightarrow \epsilon_i f_i$. Therefore, the obtained Hamiltonian describing the $f$ fermions must also enjoy the $\mathbb{Z}_2$ symmetry being consistent with \eqref{Hf}.                    

\subsection{Uncondensed phase: fractionalization and $\mathbb{Z}_2$ topological order}\label{topological_order}
In the regime $J_K<J^c_K$, the superconductor and spinons are decoupled from each other at least at the mean-field level due to vanishing $Q=0$. In order to understand this phase, we start from the charged $f$ fermions in the regime $J_K>J^c_K$ discussed in the preceding subsection. This is a reasonable starting point since we are interested in low-energy states below the original superconducting gap $\Delta$, and therefore, we only need to consider $f$ fermions.

We would like to present a rough physical picture by noting that the Kondo exchange coupling in \eqref{H_SKL} microscopically originates from the local Hubbard interaction $U$ between electrons living on the magnetic sites, resembling of Anderson impurity model. The corresponding lattice model is written as,

\begin{equation}\label{Anderson_model}
H=H_{c}[c^{\dagger},c]+\sum_{j}\left(wc^{\dagger}_{j,\alpha}f_{j,\alpha}+\mathrm{h.c}\right)+U\sum_{j}\left(n_{j,\uparrow}-\frac{1}{2} \right)\left( n_{j,\downarrow}-\frac{1}{2}\right),
\end{equation}   
where $n_{j,\alpha}=f^{\dagger}_{j,\alpha}f_{j,\alpha}$ is the local electron density on magnetic sites with spin $\alpha$ and $w$ is the overlap between electron wave functions of superconductor and magnetic sites. Note that here we denote the electron operator on the magnetic sites by $f$ keeping in mind that they are electrically charged fermions. The Kondo exchange coupling $J_{K}\simeq w^2/U$ arises in the limit of strong onsite Hubbard interaction $U\gg\Delta$ at the Anderson particle-hole symmetric point \cite{Salomaa1988}, the details are relegated to Appendix \ref{SW_transformation}. Thinking this way, the critical Kondo coupling can be associated with a critical Hubbard interaction, $J_{K}^c\simeq w^2/U_c$. Now one can identify the transition from $J_K>J^c_K$ to $J_K<J^c_K$ as a Mott transition from $U<U_c$ to $U>U_c$, respectively. Since we are interested in the low-energy states, we can start from the Bardeen–Cooper–Schrieffer (BCS) superconducting ground state of \eqref{Hf} and project it to single-occupied sites due to the Mott phase when $U>U_c$, the Gutzwiller projection. This is a way to obtain resonating valence bond (RVB) states \cite{Kotliar:prb1988} from BCS superconducting ground states.

Using the Majorana representation $f_{j,\alpha}=\left(\lambda^{[-]}_{j,\alpha}-i\lambda^{[+]}_{j,\alpha}\right)/2$, where $\lambda^{[\pm]}_{j,\alpha}$ are Majorana fermions satisfying Clifford algebra $\left\{\lambda^{[\pm]}_{j,\alpha},\lambda^{[\pm]}_{j',\alpha'}\right\}=2\delta_{jj'}\delta_{\alpha\alpha'}$, the \eqref{Hf} can be written as

\begin{equation}\label{Hf2}
H_f=-t_f\sum_{j,\alpha}\hat{U}^{\alpha}_{\mathbf{r}_j,\mathbf{r}_j+\mathbf{a}},
\end{equation}        
where $\mathbf{a}=\hat{x},\hat{y}$ is the lattice constant, and we define the bond operator $\hat{U}^{\alpha}_{\mathbf{r}_j,\mathbf{r}_j+\mathbf{a}}=i\lambda^{[+]}_{j,\alpha}\lambda^{[-]}_{j+1,\alpha}$. They form a set of commuting operators and square identity $\left(\hat{U}^{\alpha}_{\mathbf{r}_j,\mathbf{r}_l}\right)^2=1$, and hence the eigenstates can be labeled by numbers $U_{j,l}=\pm1$ on bonds. The Majorana Hamiltonian in \eqref{Hf2} has been used in Ref.[\onlinecite{Wen:PRL2003}] as a mean-field Hamiltonian to construct a $\mathbb{Z}_2$ bosonic spin liquid using the projective symmetry group \cite{Wen:PRB2003}. The latter $\mathbb{Z}_2$ symmetry is the symmetry transformation of \eqref{Hf}, whose elements form the so-called invariant gauge group. In terms of Majorana operators the single occupation constraint is given by $\lambda^{[-]}_{j,\uparrow}\lambda^{[+]}_{j,\uparrow}\lambda^{[-]}_{j,\downarrow}\lambda^{[+]}_{j,\downarrow}=1$ for all sites which determine the physical space. The operators acting within this space have to commute with the constraint, namely, they should be a product of $\hat{U}^{\alpha}_{j,l}$. The minimal operator is the fux operator $\hat{F}_{\mathbf{r}_j}=\hat{U}^{\uparrow}_{\mathbf{r}_j,\mathbf{r}_j+\hat{x}} \hat{U}^{\downarrow}_{\mathbf{r}_j+\hat{x},\mathbf{r}_j+\hat{x}+\hat{y}} \hat{U}^{\uparrow}_{\mathbf{r}_j+\hat{y},\mathbf{r}_j+\hat{x}+\hat{y}} \hat{U}^{\downarrow}_{\mathbf{r}_j,\mathbf{r}_j+\hat{y}}$ corresponding to each square plaquette, and the low-energy Hamiltonian reads as $H_{\mathrm{wp}}=16g\sum_{j}\hat{F}_{\mathbf{r}_j}$, where $g$ is an energy coupling, known as Wen plaquette model \cite{Wen:PRL2003}. Within the physical subspace, this Hamiltonian can be written in terms of spin operators. Using the Majorana representation of the Pauli operators $\sigma^{x}_{j}=i\lambda^{[+]}_{j,\downarrow}\lambda^{[-]}_{j,\uparrow}$, $\sigma^{y}_{j}=i\lambda^{[+]}_{j,\downarrow}\lambda^{[+]}_{j,\uparrow}$, $\sigma^{z}_{j}=i\lambda^{[+]}_{j,\uparrow}\lambda^{[-]}_{j,\uparrow}$, the plaquette Hamiltonian reads as $H_{\mathrm{wp}}=g\sum_{j}S^{x}_{\mathbf{r}_j} S^{y}_{\mathbf{r}_j+\hat{x}} S^{x}_{\mathbf{r}_j+\hat{x}+\hat{y}} S^{y}_{\mathbf{r}_j+\hat{y}}$, where $\mathbf{S}_j=\boldsymbol{\sigma}_j/2$, which describes a $\mathbb{Z}_2$ topologically ordered phase. Note that this is the low-energy description of the phase in the entire regime $0<J_K<J_K^c$, not just a perturbative description for $J_K\ll J_K^c$ obtained in Ref.[\onlinecite{Hsieh2017}].

\subsection{Uncondensed phase: subgap states}\label{YSR}    
\begin{figure}[t]
\center
\includegraphics[width=\linewidth]{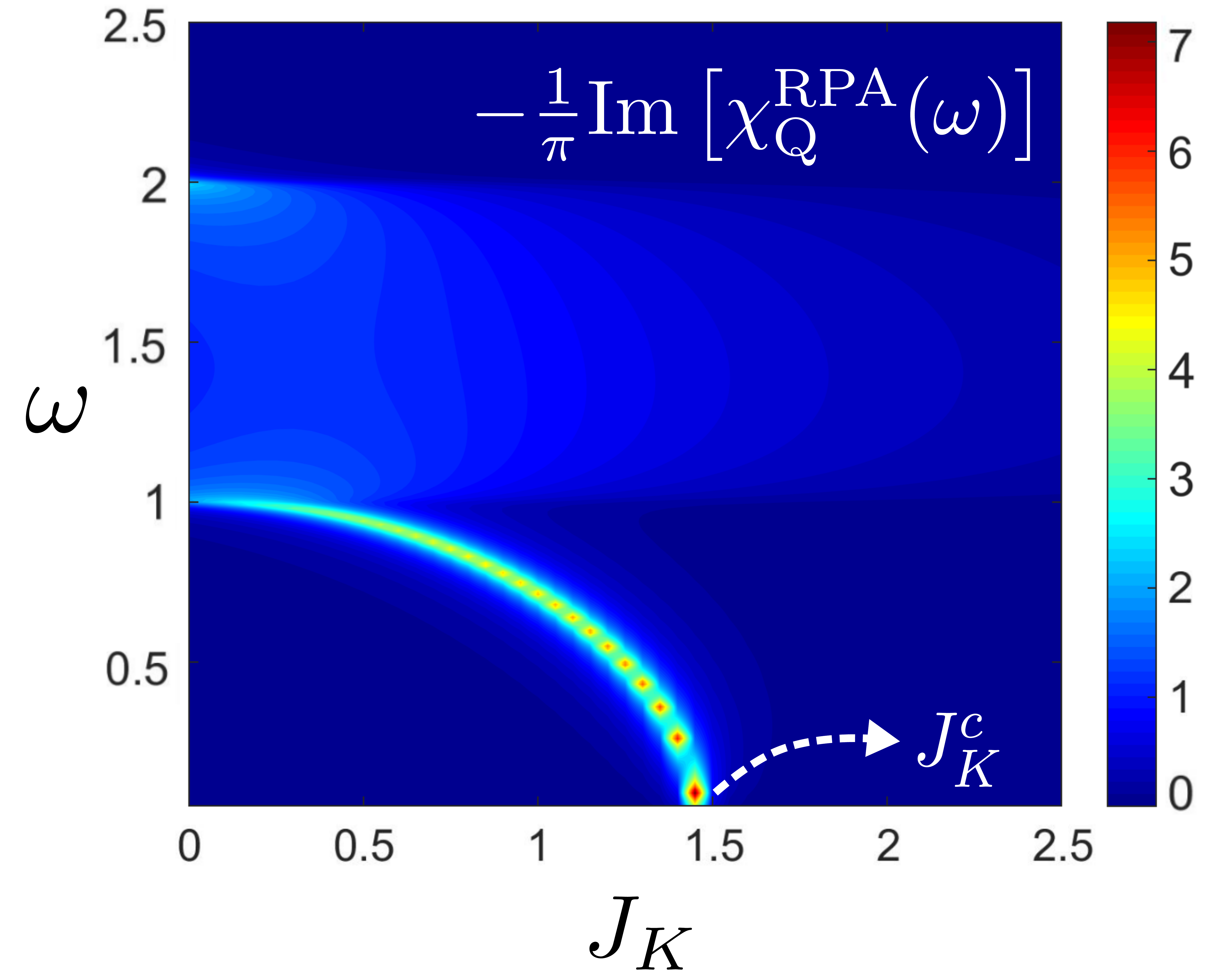}
\caption{ Spectral density $A(\omega)=(-1/\pi)\mathrm{Im}[\chi_{Q}^{\mathrm{RPA}}(\omega)]$ in \eqref{chi_RPA} evaluated for parameters $t=1$ and $\Delta=0.5$. A branch of collective excitations appears inside the superconducting gap in the region $0<\omega<2\Delta$ whose energy vanishes at the critical point $J_{K}^c$. These are the particle-hole excitations associated with the formation of subgap YSR states. The latter appears in the uncondensed phase where $Q=0$ and the magnetic moments are partially screened by Bogoliubov quasiparticles.\label{Fig:YSR}}
\end{figure}
      
In the regime $J_K<J_K^c$ the Kondo screening length $\xi_{K}$ is larger than the superconducting coherence length $\xi_{\mathrm{sc}}$. Although the superconducting gap competes with the Kondo cloud formation due to the destruction of the Fermi surface, the magnetic impurities are partially screened by superconducting quasiparticles, and bound states are formed in the superconducting gap. This is the essence of the formation of the YSR states in superconductors \cite{Yu1965, Shiba1968, Rusinov1968}. In fact, the superconductor does not destroy the Kondo cloud completely as shown for an $s$-wave superconductor\cite{Moca:PRL2021} using renormalization group analysis and numerical calculations.

The YSR states are indeed the particle-hole excitations and hence can be examined by evaluating correlation functions. The proper response function is the RPA susceptibility in \eqref{chi_RPA}. The spectral density $A(\omega)=(-1/\pi)\mathrm{Im}\left[\chi^{\mathrm{RPA}}_{Q}(\omega)\right]$ is shown in Fig.~\ref{Fig:YSR}. We set the parameters as $t=1$ and $\Delta=0.5$. The region $0<\omega<2\Delta$ corresponds to the superconducting gap, and the Bogoliubov quasiparticles have nonvanishing weight in the region above the gap, $\omega>2\Delta$, as expected. The most prominent feature of the spectral density is the appearance of a branch of subgap excitations when the Kondo coupling is less than the critical coupling, $J_K<J_K^c$. By increasing $J_K$, the energy of subgap states decreases and vanishes at the critical point. Also, we note that the weight of excitations increases as the critical point is approached, implying that more and more quasiparticles bind the magnetic moments, and the phase transition for $J_K>J_K^c$ is accompanied by the singlet formation as discussed in Sec.\ref{condensed}; a singlet is local in real space and hence includes bound states of quasiparticles of all momenta.           

\section{numerical results: quasi-one-dimensional model}\label{quasi_one_dimension}
\begin{figure}[t]
\center
\includegraphics[width=\linewidth]{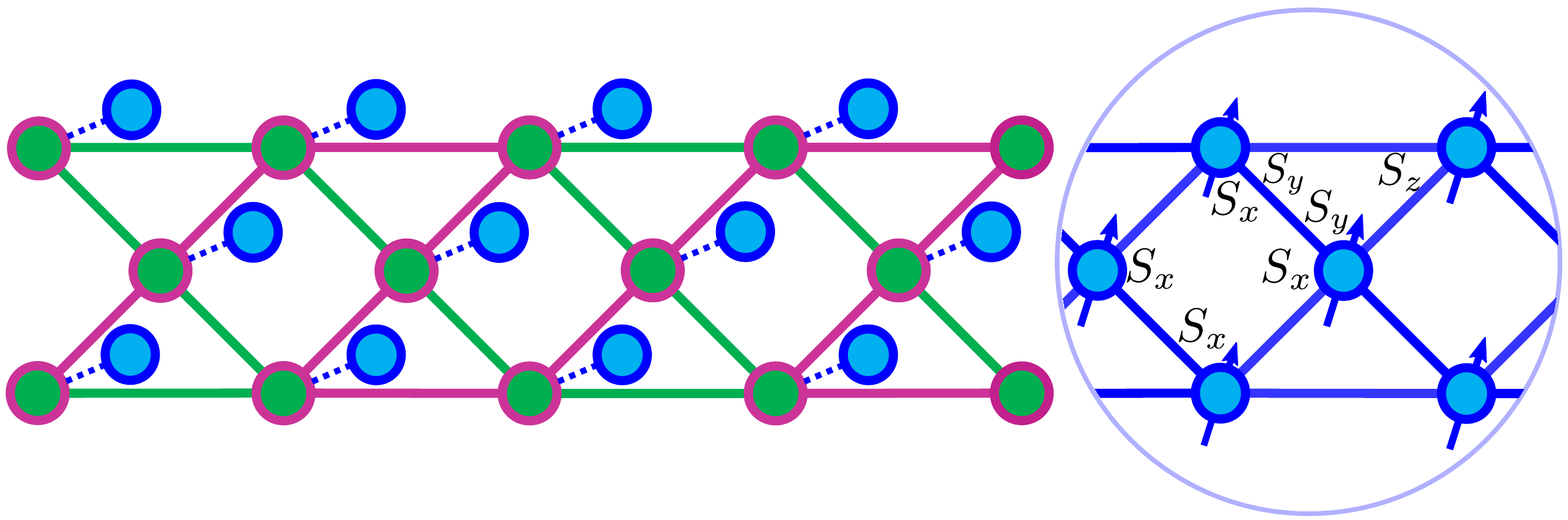}
\caption{ Left: a schematic of quasi-one-dimensional version of superconducting Kondo lattice model adapted for numerical DMRG calculations. The model consists of red (green) Kitaev chains with spin-up (down) electrons. Blue circles are magnetic moments and are Kondo coupled to the superconductor. Right: the low-energy effective model is described by a 1D version of the Wen plaquette model with plaquette operators as a product of spin operators indicated in the triangular and square plaquettes.\label{quasi_1D_toric_code_scheme}}
\end{figure}

In this section, we use the density matrix renormalization group (DMRG) to explore further the phase transition in the superconducting Kondo lattice model. For that we introduced a quasi-one-dimensional lattice shown in Fig.~\ref{quasi_1D_toric_code_scheme}. First, we use perturbation theory to derive the effective spin model with $\mathbb{Z}_2$ topological order.

The model shown in Fig.~\ref{quasi_1D_toric_code_scheme} comprises of two Kitaev chains. The red (green) chains carry spin up (down) electrons as before and is described by Majorana Hamiltonian $H_{\sigma} = \sum_{j} i \lambda^{[+]}_{j,\sigma} \lambda^{[-]}_{j+1,\sigma}$ where we set $t = \Delta = 1$ and $\mu = 0$ in \eqref{Kitaev_chain}, and each site is Kondo coupled to magnetic ions, blue circles in Fig.~\ref{quasi_1D_toric_code_scheme}. For weak coupling, the effective Hamiltonian reads as
\begin{equation}\label{H_1D}
H_{\mathrm{eff}} = - \frac{3J_K^3}{2^9} \sum_{\triangle} S^y S^z S^y  + \frac{3J_K^4}{2^{12}} \sum_{\bigDiamond} \prod_{j \in \bigDiamond} S_j^x
\end{equation}
where the first and second sums run over triangles and squares plaquettes shown in Fig.~\ref{quasi_1D_toric_code_scheme}. The terms in \eqref{H_1D} are commuting stabilizer operators, and the model possesses $\mathbb{Z}_2$ topological order. It has been recently shown that the latter topological order on a ladder geometry exhibits interesting dynamical behavior, the quasi-many body localization via self-induced disorder, in their highly excited states as well \cite{Yarloo2018}. 

\begin{figure}[t]
\center
\includegraphics[width=\linewidth]{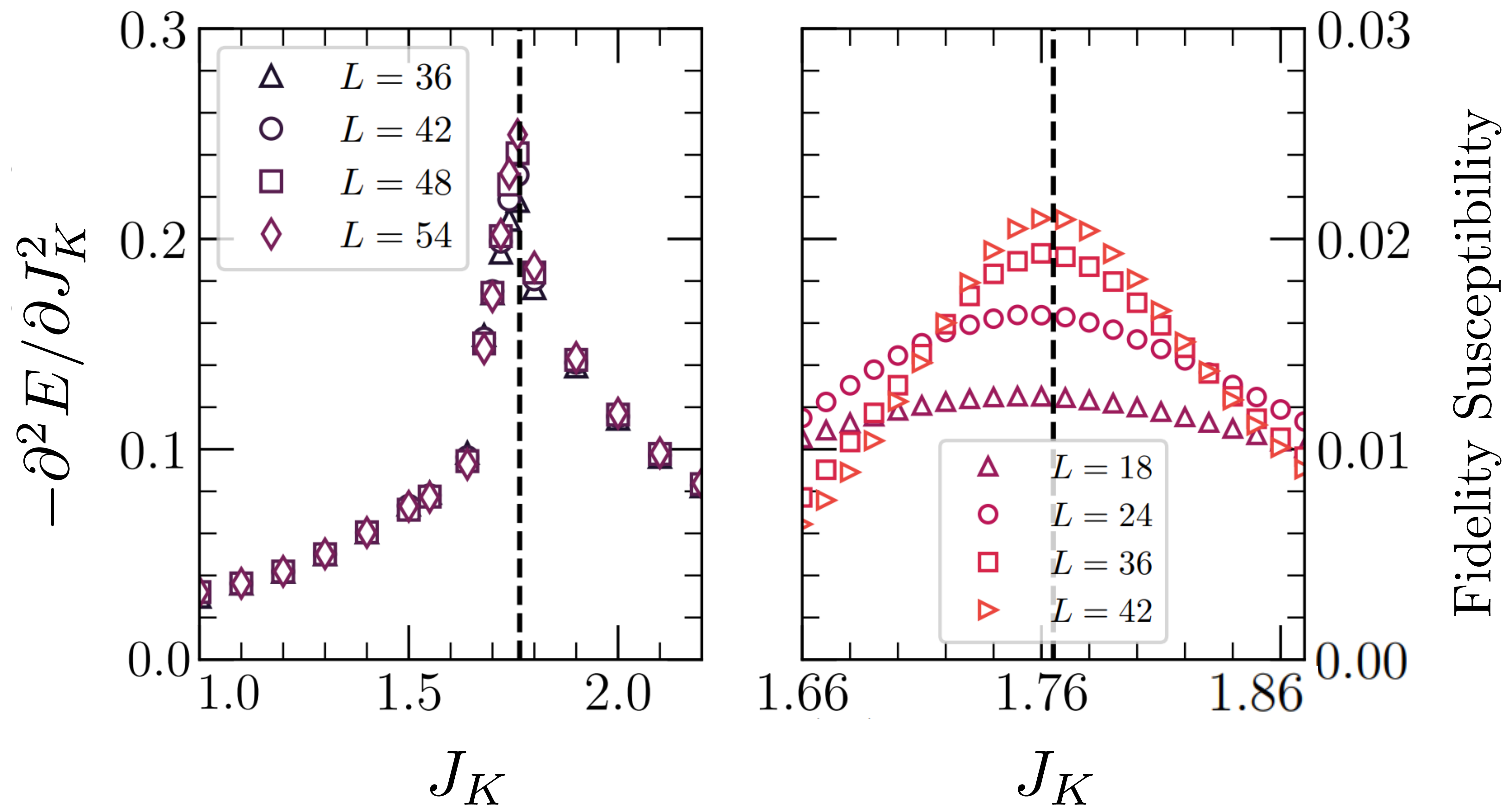}
\caption{The second derivative of the ground-state energy density and fidelity susceptibility as a function of interaction strength $J_{K}$ for various system sizes $L$. Here we set $\delta J_{K} = 0.0025$. Both quantities show a clear peak near $J_{K}\thicksim1.76$ narrowed by increasing $L$, heralding topological phase transition.\label{DMRG_phase_transition}}
\end{figure}

We proceed with the DMRG algorithm to show that the quasi-one-dimensional model undergoes a phase transition from a topologically ordered phase described by the effective Hamiltonian \eqref{H_1D} to a trivial phase. Here we use two separate $\mathbb{Z}_2$ symmetries associated with Kondo interaction defined in even and odd sublattices allowing us to increase bond dimension up to $\chi=1024$ for a large system of size $L=54$. The phase transition is identified by observing developments of singularity in the second derivative of energy density, and fidelity susceptibility \cite{Cozzini2007, Zhou2007, Zhou2008}, which measures the changes of ground state $|\psi_{0}(J_K)\rangle$ due to an infinitesimal increase in Kondo strength and is defined for a system size $L$ as
\begin{equation}\label{Susep}
\begin{split}
&\mathcal{S}(J_K)=\lim _{\delta J_K \rightarrow 0} \frac{2[1-\mathcal{F}(J_K, J_K+\delta J_K)]}{L \delta J_K^{2}}, \\
&\mathcal{F}(J_K, J_K+\delta J_K)=\left|\left\langle\psi_{0}(J_K) |\psi_{0}(J_K+\delta J_K)\right\rangle\right|
\end{split}
\end{equation}
where $\mathcal{F}(J_K, J_K+\delta J_K)$ is the ground state fidelity. As indicated in Fig~\ref{DMRG_phase_transition}, the phase transition occurs near $J_K^c \simeq 1.76$ and is narrowed by increasing system size.

\begin{figure}
\center
\includegraphics[width=\linewidth]{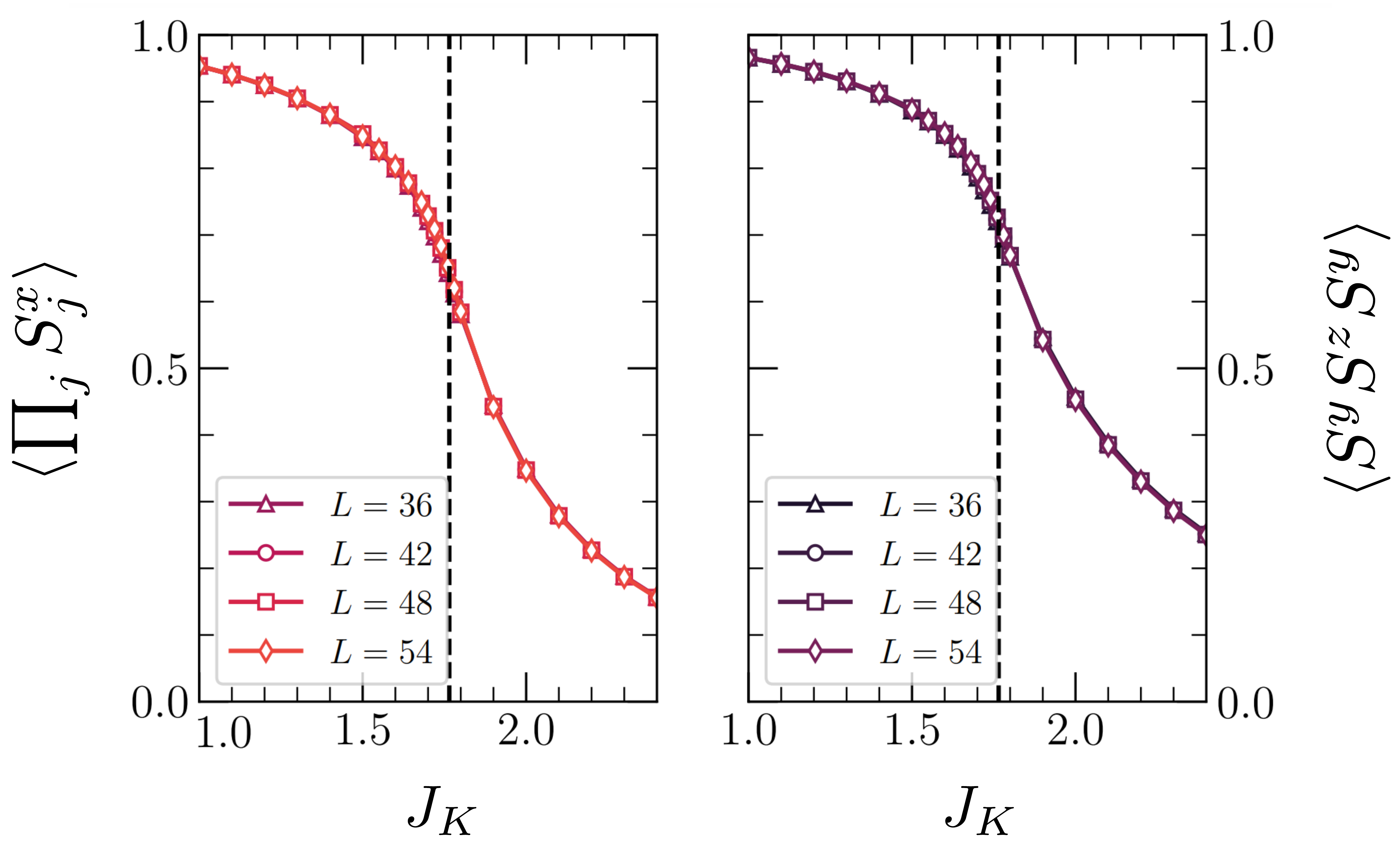}
\caption{Mean absolute value of stabilizer operators  $|\langle \prod_{j} S^x_j \rangle|$ and $\langle S^{y}S^{z}S^{y}\rangle$ as a function of interaction strength $J_{K}$ averaged over diamonds and triangles, respectively. Both plaquette operators undergo a drastic reduction near the critical point.}\label{plaquette_operators}
\end{figure}

In Fig.~\ref{plaquette_operators} we present the expectation values of plaquette stabilizer operators $|\langle \prod_{j} S^x_j \rangle|$ and $\langle S^{y}S^{z}S^{y}\rangle$. In the topologically ordered phase, they assume values close to unity. Our results show that the expectation values decrease substantially beyond $J_K^c \simeq 1.76$ in the trivial phase.

\begin{figure}[t]
\center
\includegraphics[width=\linewidth]{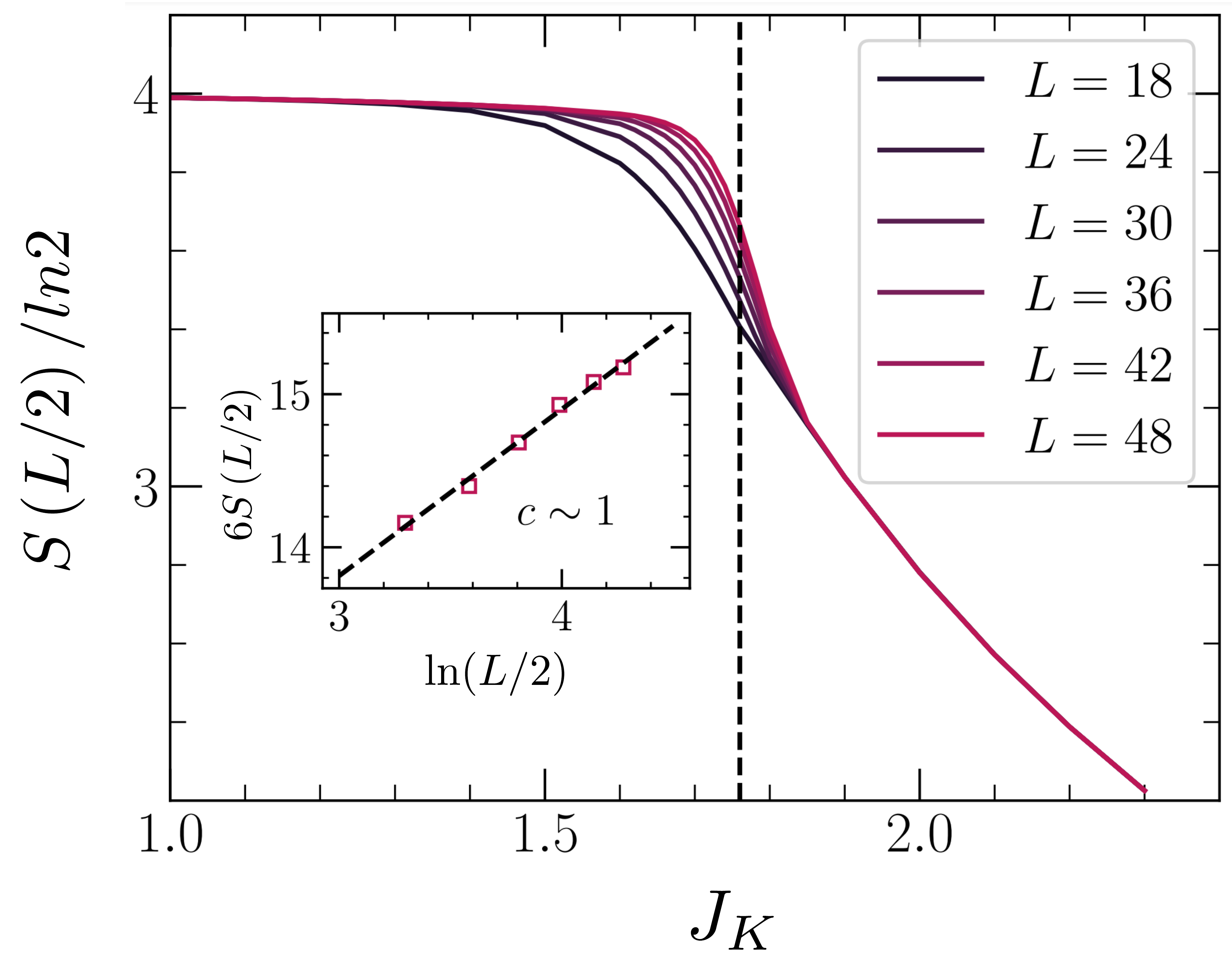}
\caption{Half-cut entanglement entropy of the system for various system sizes, $L$, as a function of interaction strength $J_{k}$. Inset: linear fit of the rescaled entanglement entropy according to Eq.~\eqref{eq:cardy}, which gives a fair approximation of the central charge $c\simeq1$. \label{central_charge}}
\end{figure} 

To get insight into the universality class of the phase transition, we evaluate the central charge, $c$, using the scaling of entanglement entropy near the critical point by the Cardy-Calabrese formula \cite{Holzhey1994, Vidal2003, Calabrese2009}

\begin{equation}  \label{eq:cardy}
S(x)=\frac{c}{6} \ln\left[\frac{N}{\pi} \sin \left(\pi \frac{x}{N}\right)\right]+\text {const. },
\end{equation}
where $N$ is the length of lattice and $x$ denotes the cut length.
As shown in Fig.~\ref{central_charge}, the half-cut entanglement entropy exhibits a plateau below $J_{K}^c$ strictly at $S_{L/2}=4\ln 2$, which tends to zero as expected from the topologically trivial phase. The scaling of $S_{L/2}$ near the critical point, see the inset of Fig.~\ref{central_charge}, displays a logarithmic correction of area law entanglement, so the central charge $c\simeq1$ is twice that of the transverse field Ising (TFI) chain, i.e., $c_{\mathrm{TFI}}=1/2$.

We also numerically verified that at the quantum critical point, $J^c_K \simeq 1.76$, the two-point correlation functions for $c$ and spin operators $\mathbf{S}$ are both decaying algebraically. The fact that $c=1$ at the quantum critical point implies that a combination of $c$ fermions forms a single critical (gapless) Majorana chain (in addition to other potentially massive branches), thus contributing $c=1/2$ to the observed central charge. The gaplessness of $c$ fermions at the critical point can be readily seen by examining the poles of the Green function $[\mathcal{G}_{c,\alpha}(\mathbf{k},ik_n)]^{-1}=[\mathcal{G}^0_{c,\alpha}(\mathbf{k},ik_n)]^{-1}-\Sigma_{c,\alpha}(ik_n)$ after analytical continuation $ik_n\to \omega^{+} (\omega^{+}=\omega+i0^{+})$, where $\Sigma_{c,\alpha}(ik_n)$ is the self-energy of $c$ electrons coupled to $f$ fermions via the renormalized interaction $V(iq_n)=V_0/(1-V_{0}\chi_{Q}^{0}(iq_n))$ (see \eqref{chi_RPA}):

\begin{equation}
\mathcal{G}^0_{c,\alpha}(\mathbf{k},\omega^{+})=\left[\omega^{+} - \varepsilon(k_{\alpha}) - \frac{|\Delta(k_{\alpha})|^2}{\omega^{+} + \varepsilon(k_{\alpha})}\right]^{-1},
\end{equation}

\begin{equation}
\Sigma_{c,\alpha}(\omega^+)=-P\int_{-\infty}^{\infty}\frac{d\Omega}{\pi}\frac{V_0n_B(\Omega)}{\omega^{+}-\Omega}\mathrm{Im}\left[\frac{1}{\varepsilon(\Omega^{+})} \right]+\frac{V_0}{2\varepsilon(\omega^{+})},
\end{equation}   
where $n_B(\Omega)$ is the bosonic distribution function, and $\varepsilon(\Omega^{+})=1-V_{0}\chi_{Q}^{0}(\Omega^{+})$ with $\Omega^{+}=\Omega+i0^{+}$ is the RPA dielectric constant in \eqref{chi_RPA}. The $P$ in the first term stands for principle value. Furthermore, the $f$ spinons must form another single critical Majorana chain responsible for the remaining $c=1/2$ in the measured central charge. Note that the obtained central charge is consistent with the quantum dimension, $d=\sqrt{2}$, of a Majorana mode localized on each edge through $c=\log_2 d$ \cite{Verresen2017}.

As $J_K \to 0$, the $c$ fermions are gapped and the perturbation theory is reliable and valid. As we start to increase $J_K$ two phenomena occur concurrently: (i) the excitation gap for the $c$-fermions starts to diminish and eventually vanishes at $J^c_K\simeq 1.76$. Therefore, our perturbation expansions which assumed a non-zero excitation gap for the Kitaev's chains, will fail beyond critical point and as a result, the emergent $\mathbb{Z}_2$ spin liquid phase will no longer exist for $J_K \geq 1.76$. (ii) As argued before, the $f$-spinons have a non-zero Mott gap for $J_K < 1.76$, which justifies the projection of the quadratic terms in \eqref{Hf} to single  occupied sites subspace, which eventually leads to the Kitaev's toric code model as the effective Hamiltonian. At the critical coupling, $J_K \approx 1.76$, both the $c$-fermions and $f$-spinons become gapless, each with a $c=1/2$ central charge conformal field theory description.

\section{CONCLUSIONS}\label{conclusions}
In this work, we studied the topological phases and phase transition in a superconducting Kondo lattice model, a lattice of topological superconductors is antiferromagnetically coupled to otherwise free magnetic ions near each site. Using the slave-fermion representation of spins and a proper Hubbard-Stratonovich decomposition, the model becomes a superconducting lattice of electrons coupled to spinons through a fluctuating bosonic field. The mean-field theory shows that the bosonic field, which itself is a hybridization between superconducting electrons and spinons, condenses beyond a finite critical Kondo exchange coupling. This is in contrast with the metallic Kondo lattice model, where the Kondo coupling is marginally relevant. 

Our results show that the condensed phase is an invertible topological phase, i.e., the spinons become a topological superconductor characterized by an invariant with the opposite sign of the superconducting electrons. Since the invariant of the electrons and spinons vanishes together, the total system is trivial, which is consistent with singlet formation at the strong coupling limit. The phase transition to the uncondensed phase by reducing the Kondo coupling can be interpreted as a Mott transition for the charged spinons, and the corresponding low-energy phase is topologically ordered with underlying $\mathbb{Z}_2$ gauge theory. Further, we showed that the subgap states are formed in the uncondensed phase due to the Kondo screening of magnetic moments by superconducting quasiparticles. 

To further corroborate the existence of a phase transition in the superconducting Kondo lattice model, we used numerical DMRG on a quasi-one-dimensional lattice of the model. The numerical results indicate that the phase transition has a topological nature, with plaquette operators acquiring values close to unity in the topologically ordered phase and approaching zero in the phase with no topological order. We also used the numerics on finite-size lattices to determine the critical nature of the phase transition by evaluating the central charge.  

Before closing, we would like sketch a few directions which could be interesting for future studies. The work presented here along the works in Refs.[\onlinecite{Hsieh2016, Hsieh2017, Mohammadi2022}] may provide a conceptual framework to design more exotic topological orders such as non-Abelian anyons. In particular, the physical pictures presented may pave the way to look for proper materials, most likely in layered materials or heterostructures, with approperiate coupled electronic degrees of freedom. Also, our mean-field study could be further enriched by assuming a nonuniform ansatz. The hopping field, $Q_j$, in \eqref{Hcf} could carry non-trvial mometum, such as in the staggered flux phase. In this work, for the sake of simplicity, we ignored those more complex structures. The impact of those possibilities can be an interesting future direction.

\appendix
\begin{widetext}
\section{The superconducting Green's function } \label{Green_SC}

The finite-temperature superconducting field operator $\hat{\vec{c}}_{\vec{k}\sigma}(\tau) =\left(\begin{matrix} c_{\vec{k}, \sigma}(\tau) & c_{-\vec{k}, \sigma}^{\dagger}(\tau)\end{matrix}\right)^T$ evolves by the mean-filed Hamiltonian \eqref{H_MF}. We define Green's function matrix of superconducting subspace as follows
\begin{equation}
\mathbf{G}_{\sigma}(\vec{k}, \tau) = -\langle T_{\tau} [\vec{c}_{\vec{k}\sigma}(0) \ \vec{c}_{\vec{k}\sigma}^{\dagger}(\tau)] \rangle 
 = \left(\begin{matrix}
\mathcal{G}_{c,\sigma}(\vec{k}, \tau) & \mathcal{F}_{c, \sigma}(\vec{k}, \tau) \\
\mathcal{F}^{\dagger}_{c, \sigma}(\vec{k}, \tau) & -\mathcal{G}_{c,\sigma}(\vec{k}, -\tau)
\end{matrix}\right),
\end{equation}
where $\mathcal{G}_{c,\sigma}(\vec{k}, \tau)$ and $\mathcal{F}_{c, \sigma}(\vec{k}, \tau)$ are single-particle and anomalous Green's functions, respectively. Differentiating $\mathbf{G}$ with respect to $\tau$ yields
\begin{equation}
\begin{split}
&\partial_{\tau} \mathcal{G}_{c, \sigma}(\vec{k}, \tau) = - \Delta^*(k_{\sigma}) \mathcal{F}^{\dagger}_{c, \sigma}(\vec{k}, \tau) + Q^* \mathcal{G}_{cf, \sigma}(\vec{k}, \tau)  -\varepsilon(k_{\sigma}) \mathcal{G}_{c, \sigma}(\vec{k}, \tau) -\delta(\tau),\\
&\partial_{\tau} \mathcal{F}^{\dagger}_{c, \sigma}(\vec{k}, \tau) =  \varepsilon(k_{\sigma}) \mathcal{F}^{\dagger}_{c, \sigma}(\vec{k}, \tau) - \Delta(k_{\sigma}) \mathcal{G}_{c, \sigma}(\vec{k}, \tau) -Q \mathcal{F}^{\dagger}_{cf, \sigma}(\vec{k}, \tau).
\end{split}
\end{equation}
This led us to consider the new Green's functions $\mathcal{G}_{cf, \sigma}(\vec{k}, \tau) = -\langle T_{\tau} [f_{\vec{k}, \sigma}(\tau) c_{\vec{k}, \sigma}^{\dagger}(0)]\rangle$ and $\mathcal{F}^{\dagger}_{cf, \sigma}(\vec{k}, \tau) = -\langle T_{\tau} [f^{\dagger}_{-\vec{k}, \sigma}(\tau) c_{\vec{k}, \sigma}^{\dagger}(0)] \rangle$ with imaginary time derivatives
\begin{equation}
\begin{split}
&\partial_{\tau} \mathcal{G}_{cf, \sigma}(\vec{k}, \tau) = Q \mathcal{G}_{c, \sigma}(\vec{k}, \tau) - \lambda \mathcal{G}_{cf, \sigma}(\vec{k}, \tau) \\
& \partial_{\tau} \mathcal{F}^{\dagger}_{cf, \sigma}(\vec{k}, \tau) =  -Q^* \mathcal{F}^{\dagger}_{c, \sigma}(\tau) + \lambda \mathcal{F}^{\dagger}_{cf, \sigma}(\vec{k}, \tau)
\end{split}
\end{equation}
The corresponding equations of motion in the Matsubara domain read as
\begin{equation}\label{Matsubara_domain}
\begin{split}
&\left( -ik_n + \varepsilon(k_{\sigma})\right) \mathcal{G}_{c, \sigma}(\vec{k}, ik_n) =  -\Delta^*(k_{\sigma})  \mathcal{F}^{\dagger}_{c, \sigma}(\vec{k}, ik_n) + Q^*  \mathcal{G}_{cf, \sigma}(\vec{k}, ik_n) -1,\\
&\left(-ik_n - \varepsilon(k_{\sigma}) \right) \mathcal{F}^{\dagger}_{c, \sigma}(\vec{k}, ik_n) = -\Delta(k_{\sigma})  \mathcal{G}_{c, \sigma}(\vec{k}, ik_n) - Q  \mathcal{F}^{\dagger}_{cf, \sigma}(\vec{k}, ik_n),\\
&-ik_n  \mathcal{G}_{cf, \sigma}(\vec{k}, ik_n) = Q  \mathcal{G}_{c, \sigma}(\vec{k}, ik_n) - \lambda \mathcal{G}_{cf, \sigma}(\vec{k}, ik_n),\\
&-ik_n  \mathcal{F}^{\dagger}_{cf, \sigma}(\vec{k}, ik_n) = -Q  \mathcal{F}^{\dagger}_{c, \sigma}(\vec{k}, ik_n) + \lambda  \mathcal{F}^{\dagger}_{cf, \sigma}(\vec{k}, ik_n),
\end{split}
\end{equation}
where $k_n = (2n+1) \pi T$ is Matsubara frequencies in temperature $T$. These equations are readily solved to give
\begin{equation}\label{g_cf}
\begin{split}
&\mathcal{G}_{c, \sigma}(\vec{k}, ik_n) = \left[ik_n - \varepsilon(k_{\sigma}) - \frac{|Q|^2}{ik_n - \lambda} - \frac{|\Delta(k_{\sigma})|^2}{ik_n + \varepsilon(k_{\sigma}) - \frac{|Q|^2}{ik_n + \lambda}}\right]^{-1},\\
&\mathcal{G}_{cf, \sigma}(\vec{k}, ik_n) = \frac{-Q}{ik_n - \lambda} \mathcal{G}_{c, \sigma}(\vec{k}, ik_n).
\end{split}
\end{equation}
Following the same procedure lead to \eqref{g_cf}, we differentiate spinons single-particle Green's function, i.e. $\mathcal{G}_{f, \sigma}(\vec{k}, \tau) = -\langle T_{\tau} [f_{\vec{k}, \sigma}(\tau)f_{\vec{k}, \sigma}^{\dagger}(0)]\rangle$, and other corresponding Green's functions with respect to $\tau$ in Matsubara domain
\begin{equation}\label{g_f_eqs}
\begin{split}
&(-ik_n + \lambda) \mathcal{G}_{f, \sigma}(\vec{k}, ik_n) = Q \mathcal{G}_{fc, \sigma}(\vec{k}, ik_n) -1, \\
&(-ik_n + \varepsilon(k_{\sigma})) \mathcal{G}_{fc, \sigma}(\vec{k}, ik_n) = -\Delta^*(k_{\sigma}) \mathcal{F}^{\dagger}_{fc, \sigma}(\vec{k}, ik_n) + Q^* \mathcal{G}_{f, \sigma}(\vec{k}, ik_n), \\
&(-ik_n - \varepsilon(k_{\sigma})) \mathcal{F}^{\dagger}_{fc, \sigma}(\vec{k}, ik_n) = -\Delta(k_{\sigma}) \mathcal{G}_{fc, \sigma}(\vec{k}, ik_n) - Q \mathcal{F}^{\dagger}_{f, \sigma}(\vec{k}, ik_n), \\
&(-ik_n - \lambda) \mathcal{F}^{\dagger}_{f, \sigma}(\vec{k}, ik_n) = -Q^* \mathcal{F}^{\dagger}_{fc, \sigma}(\vec{k}, ik_n),
\end{split}
\end{equation} 
which determine the Green's function of spinon as
\begin{equation}\label{A_Gf}
\mathcal{G}_{f, \sigma}^{-1}(\vec{k}, ik_n) = ik_n - \lambda - \frac{|Q|^2}{ik_n - \varepsilon(k_{\sigma}) - \frac{|\Delta(k_{\sigma})|^2}{ik_n + \varepsilon(k_{\sigma}) - \frac{|Q|^2}{ik_n + \lambda}}}.
\end{equation}

\section{The massive gauge fluctuations } \label{Higgs}
\subsection{phenomenological theory}
Under the U(1) gauge transformation $f_{j}\rightarrow e^{i\phi_{j}(\tau)}f_{j}$, the hybridization field transforms as $Q_{j}\rightarrow e^{i\phi_{j}(\tau)}Q_j$ and the constraint field transforms as $\lambda_{j}\rightarrow\lambda_{j}-\partial_{\tau}\phi_{j}(\tau)$. Let us define a gauge field $A_{ij}$ living on the bond so that under the gauge transformation $A_{ij}\rightarrow A_{ij}+\phi_{j}(\tau)-\phi_{i}(\tau)$. Thus $\lambda$ and $A$ can be imagined as temporal and spatial components of an emergent U(1) gauge potential $(A_0,\mathbf{A})$. Taking the continuum limit $\phi_{j}(\tau)\rightarrow \phi(\mathbf{r},\tau)$ , the low-energy gauge invariant theory reads as 

\begin{align}\label{Lagrangian}
\mathcal{L}[Q,A_{\mu}]=K_1F_{\mu\nu}F^{\mu\nu}+K_2\left|(i\partial_{\mu}-A_{\mu})Q \right|^2+r|Q|^2+u|Q|^4+\cdots,
\end{align}
where the first term $F_{\mu\nu}=\partial_{\mu}A_{\nu}-\partial_{\nu}A_{\mu}$ is the usual Maxwell term and $K_1, K_2, r, u$ are phenomenological parameters with $K_1, K_2, u>0$. The second term clearly shows that the bosons with charge unity are coupled to an emergent gauge potential. In the condensed phase $r<0~(J_{K}>J_{K}^c)$ and $Q$ develops finite values. 
The above theory is manifestly invariant under the $U(1)$ gauge transformation,

\begin{align}
Q\rightarrow e^{i\phi(\mathbf{r},\tau)}Q,~~\mathbf{A}\rightarrow\mathbf{A}-\boldsymbol{\nabla}\phi(\mathbf{r},\tau), ~~~A_0\rightarrow A_0-\partial_{\tau}\phi(\mathbf{r},\tau).
\end{align}       

In the condensed phase $Q=|Q|e^{-i\theta}$. We focus on the phase fluctuations and and ignore the amplitude mode, i.e., $|Q|=Q_0$ is constant. We obtain,

\begin{align}
\mathcal{L}[A_{0},\mathbf{A}]=K_1F_{\mu\nu}F^{\mu\nu}+K_2Q_{0}^2\left|(\partial_{\mu}\theta-A_{\mu})\right|^2
\end{align}

The $\partial_{\mu}\theta$ in the second term can be eliminated by a gauge transformation making $Q$$(=Q_0)$ to be real, i.e., by setting $\theta=\phi$ and $\tilde{A}_{\mu}=A_{\mu}-\partial_{\mu}\phi(\mathbf{r},\tau)$, and we obtain a pure U(1) gauge theory
\begin{align}
\mathcal{L}[\tilde{A}_{0},\tilde{\mathbf{A}}]=K_1F_{\mu\nu}F^{\mu\nu}+K_2Q_{0}^2\left|\tilde{A}_{\mu}\right|^2,
\end{align} 
which is manifestly massive due to second term with condensation $Q_{0}\neq0$. Note that $F_{\mu}=\tilde{F}_{\mu}$. The above expression is the well-known Proca model in (2+1)-d QFT. Thus, the phase fluctuations of the condensate are eliminated from the low-energy theory and the gauge fields become massive. This is the Higgs mechanism.

\subsection{microscopic derivation}
 Let us sketch the main lines 
of derivation of the Lagrangian density in \eqref{Lagrangian}. For sake of simplicity we only consider the dynamics of the fields and ignore the spatial variations. The effective action is derived by integrating over the fermionic fields:           

\begin{align}
e^{-\mathcal{S}[Q,\lambda]}=\int D[c,f] e^{-\mathcal{S}[Q, \lambda, c,f]},
\end{align} 
where $\mathcal{S}[Q, \lambda, c,f]$ is given in \eqref{action}. In frequency-momentum space, 

\begin{align}
\mathcal{S}[Q, \lambda, c,f]&=\sum_{k}\left(-ik_{n}\left(c^{\dagger}_{k,\alpha}c_{k,\alpha}+f^{\dagger}_{k,\alpha}f_{k,\alpha}\right)+H_{sc}[c^{\dagger},c] \right)\\
&+\sum_{q}\left(-\frac{1}{V_0}|Q(q)|^2-Q(q)\sum_{k}f^{\dagger}_{k+q,\alpha}c_{k,\alpha}-\mathrm{H.c.}  \right)-i\sum_{k,q}\lambda(q)f^{\dagger}_{k+q,\alpha}f_{k,\alpha},
\end{align} 
where $q=(iq_n,\mathbf{q})$ and $k=(ik_n,\mathbf{k})$, and the superconducting Hamiltonian $H_{sc}$ is given in \eqref{Hsc}. Integrating out the fermions along with proper contractions of $c$ and $f$ fermions lead to evaluations of polarization functions in lowest orders. We obtain,  

\begin{align}
\mathcal{S}[Q, \lambda]=\sum_{q}\left(-\frac{1}{V_0}+\chi_{Q}^{0}(iq_n)\right)|Q(q)|^2+ \sum_{q}\chi_{f}(q)|\lambda(q)|^2+\cdots
\end{align}
where
\begin{align}
\chi_{f}(q)=T\sum_{k}\mathcal{G}_{f,\alpha}(k)\mathcal{G}_{f,\alpha}(k+q)
\end{align}
with $\mathcal{G}_{f,\alpha}(k)$ and $\chi_{Q}^{0}(q)$ are given in \eqref{A_Gf} and \eqref{chi0}, respectively. Taking the low-frequency limit, they read as

\begin{align}
&\chi_{Q}^{0}(iq_n)\approx \chi_{Q}^{0}(0)-a (iq_n)^2+\cdots\\
& \chi_{f}(iq_n)\approx b+ c(iq_n)^2,
\end{align}  
where $a, b, c$ $(b, c\propto |Q|^2)$ are constants. Using \eqref{chi0_F} and \eqref{Jcritical} we define $r=2\left(J^{-1}_{K}- J^{c^{-1}}_{K}\right)$ so that $r<0$ $(J_{K}>J_{K}^c)$ corresponds to the condensed phase. Fourier transformed back to imaginary time domain and read space, we obtain

\begin{align} 
&\mathcal{S}[Q, A_0]=\int d\tau d^2x ~\mathcal{L}[Q, A_0],\\
&\mathcal{L}[Q, A_0]=K_1|\partial_{\tau}A_0|^2+K_{2}|\partial_{\tau}Q|^2+r|Q|^2+\cdots, 
\end{align}
where $A_0\equiv\lambda$ is identified.

\section{Deriving the Kondo exchange coupling} \label{SW_transformation}

In this section, we derive the Kondo exchange term in equation \eqref{H_SKL} using the Schrieffer-Wolf transformation (SW) applied to the periodic Anderson model \eqref{Anderson_model}. For the sake of simplicity we start from a single impurity orbital hybridized with the supercondutor. The modes desribes by $H = H_0 + H_w$:  

\begin{equation}
\begin{split}
&H_0 = H_{\mathrm{sc}}[c^{\dagger},c] + \varepsilon_d (n_{\uparrow}+n_{\downarrow}) + U n_{\uparrow}n_{\downarrow}\\
&H_w = w\sum_{\vec{k}, \alpha} (c_{\vec{k},\alpha}^{\dagger}d_{\alpha}+ \mathrm{H.c.}),
\end{split}
\end{equation}
with $\varepsilon_d<0$ as single orbital energy. The SW transformation unitarily  transforms $H$ to $H'=e^SHe^{-S}$ with $S^{\dagger}=-S$, from which the high-energy single orbital states, i.e., empty and double occupations, are properly eliminated as

\begin{equation}
\begin{split}
&H_{\mathrm{eff}} = P_0 H' P_0, \\
&P_0 =  \left(|\Uparrow \rangle \langle \Uparrow| + |\Downarrow \rangle \langle \Downarrow|\right) \otimes |\psi_{BCS}\rangle \langle \psi_{BCS}|,
\end{split}
\end{equation}
where $P_0$ projects to the single-occupied orbital subspace and $|\psi_{BCS}\rangle$ is the superconducting ground state. The transformed Hamiltonian is written as

\begin{equation}\label{H'}
H' = H_0 + H_w  + [S, (H_0 + H_w)] + \frac{1}{2} [S, [S, (H_0 + H_w)]]  + ... .
\end{equation}

We follow the standard procedure in SW transformation by finding $S$ such that $[S, H_0 ] =- H_w$. The transformed Hamiltonian then reads as $H' = H_0 + \frac{1}{2}[S,H_w] + O(H_w^3)$. Using the Bogoliubov transformation $c_{\vec{k},\alpha} = u^*(k_{\alpha}) \gamma_{\vec{k},\alpha} + v(k_{\alpha}) \gamma_{-\vec{k},\alpha}^{\dagger}$ with superconducting coherence factors $u(k_{\alpha}) = \cos(\theta_{k_{\alpha}}/2)e^{i\phi_{k_{\alpha}}}$ and $v(k_{\alpha}) =\sin(\theta_{k_{\alpha}}/2)$, where $\cos(\theta_{k_{\alpha}}) = \varepsilon(k_{\alpha})/E(k_{\alpha})$ and $\tan(\phi_{k_{\alpha}}) = \Im\Delta(k_{\alpha})/\Re\Delta(k_{\alpha})$, and $E(k_{\alpha}) = \sqrt{\Delta(k_{\alpha})^2+\varepsilon(k_{\alpha})^2}$, we obtain $S = S_0 + T$ as 
 
\begin{equation}
\begin{split}
&S_0 = \sum_{\vec{k},\alpha} \Big\lbrace A(k_{\alpha})u(k_{\alpha})\left(\gamma_{\vec{k},\alpha}^{\dagger}d_{\alpha}+d_{\alpha}^{\dagger}\gamma_{\vec{k},\alpha}\right)+B(k_{\alpha})v(k_{\alpha})\left(\gamma_{\vec{-k},\alpha}d_{\alpha}+\gamma_{-\vec{k},\alpha}^{\dagger} d_{\alpha}^{\dagger}\right)\Big\rbrace\\
&T = \sum_{\vec{k},\alpha} \Big\lbrace F(k_{\alpha})u(k_{\alpha})\left(\gamma_{\vec{k},\alpha}^{\dagger}d_{\alpha}+d_{\alpha}^{\dagger}\gamma_{\vec{k},\alpha}\right)+D(k_{\alpha})v(k_{\alpha})\left(\gamma_{\vec{-k},\alpha}d_{\alpha}+\gamma_{-\vec{k},\alpha}^{\dagger} d_{\alpha}^{\dagger}\right)\Big\rbrace n_{-\alpha},
\end{split}
\end{equation}
where

\begin{equation}
\begin{split}
&A(k_{\alpha}) = \frac{w}{E(k_{\alpha})-\varepsilon_d}, \quad
B(k_{\alpha}) = -\frac{w}{E(k_{\alpha})+\varepsilon_d},\\
&F(k_{\alpha}) = w\left(\frac{1}{E(k_{\alpha})-\varepsilon_d-U} - \frac{1}{E(k_{\alpha})-\varepsilon_d}\right),\\
&D(k_{\alpha}) = w\left(\frac{1}{E(k_{\alpha})+\varepsilon_d} -  \frac{1}{E(k_{\alpha})+\varepsilon_d+U}\right)
\end{split}
\end{equation}

The transformed Hamiltonian \eqref{H'} can be written to second order as 

\begin{equation}
H' =  H_0 + H_{K}+ H_{\mathrm{pairing-correlation}} + H_{\mathrm{charge}} + H_{\mathrm{direct}}+ H'_0. 
\end{equation}

Various terms are Kondo Hamiltonian

\begin{equation}\label{H_KAppen}
\begin{split}
H_{K} = -\frac{1}{2}\sum_{\vec{k},\vec{k'}} \Bigg\lbrace \frac{1}{2}\left(J_{k_{x},k'_{x}} c_{\vec{k'},\uparrow}^{\dagger}c_{\vec{k},\uparrow}-J_{k_{y},k'_{y}} c_{\vec{k'},\downarrow}^{\dagger}c_{\vec{k},\downarrow}\right) \left(d_{\uparrow}^{\dagger}d_{\uparrow} - d_{\downarrow}^{\dagger}d_{\downarrow} \right)
+ \sum_{\alpha} J_{k_{-\alpha},k'_{\alpha}} c_{\vec{k'},\alpha}^{\dagger} c_{\vec{k},-\alpha} d_{-\alpha}^{\dagger}d_{\alpha} \Bigg\rbrace
\end{split}
\end{equation}
with exchange coupling

\begin{equation}
\begin{split}
J_{k_{\alpha},k'_{\beta}} &= w \left(D(k_{\alpha})v^2(k_{\alpha}) + F(k_{\alpha}) |u(k_{\alpha})|^2 + D(k'_{\beta})v^2(k'_{\beta}) + F(k'_{\beta}) |u(k'_{\beta})|^2 \right) \\
& = w^2 \Bigg(\frac{v^2(k_{\alpha})}{E(k_{\alpha})+\varepsilon_d} -  \frac{v^2(k_{\alpha})}{E(k_{\alpha})+\varepsilon_d+U} + \frac{|u(k_{\alpha})|^2}{E(k_{\alpha})-\varepsilon_d-U} - \frac{|u(k_{\alpha})|^2}{E(k_{\alpha})-\varepsilon_d} \\&+\frac{v^2(k'_{\beta})}{E(k'_{\beta})+\varepsilon_d} -  \frac{v^2(k'_{\beta})}{E(k'_{\beta})+\varepsilon_d+U} + \frac{|u(k'_{\beta})|^2}{E(k'_{\beta})-\varepsilon_d-U} - \frac{|u(k'_{\beta})|^2}{E(k'_{\beta})-\varepsilon_d} \Bigg),
\end{split}
\end{equation}
$H_{\mathrm{pairing-correlation}}$ modifying the pairing correlation of $c$ electrons

\begin{equation}
H_{\mathrm{pairing-correlation}} =  -\frac{1}{2}\sum_{\vec{k},\vec{k'}} \Bigg\lbrace \left(\frac{1}{2}\left(W_{k'_{x}} c_{-\vec{k'},\uparrow}c_{\vec{k},\uparrow}-W_{k'_{y}} c_{\vec{-k'},\downarrow}c_{\vec{k},\downarrow}\right) \left(d_{\uparrow}^{\dagger}d_{\uparrow} - d_{\downarrow}^{\dagger}d_{\downarrow} \right) + \sum_{\alpha} W_{k'_{\alpha}} c_{-\vec{k'},\alpha} c_{\vec{k},-\alpha} d_{-\alpha}^{\dagger}d_{\alpha}\right) + \mathrm{H.c.} \Bigg\rbrace
\end{equation}
with
\begin{equation}
\begin{split}
W_{k_{\alpha}} = w^2u(k_{\alpha})v(k_{\alpha}) \left( \frac{1}{E(k_{\alpha})+\varepsilon_d+U}-\frac{1}{E(k_{\alpha})+\varepsilon_d} + \frac{1}{E(k_{\alpha})-\varepsilon_d-U} - \frac{1}{E(k_{\alpha})-\varepsilon_d}\right),
\end{split}
\end{equation}
 the charge Hamiltonian

\begin{equation}
H_{\mathrm{charge}} = \frac{1}{2} \sum_{\vec{k},\vec{k'},\alpha} \Bigg\lbrace \left(\frac{1}{2}J_{k'_{\alpha},k'_{\alpha}} c_{\vec{k'},\alpha}^{\dagger}c_{\vec{k},-\alpha}^{\dagger} +  W_{k'_{\alpha}}c_{-\vec{k'},\alpha}c_{\vec{k},-\alpha}^{\dagger}\right) d_{\alpha} d_{-\alpha} + \mathrm{H.c.} \Bigg\rbrace,
\end{equation}

which changes the $d$ orbital occupation by two electrons, the direct spin-independent interaction
\begin{equation}
\begin{split}
H_{\mathrm{direct}} = \frac{1}{2}\sum_{\vec{k},\vec{k'},\alpha}\left\lbrace\left(\frac{1}{2}J_{k_{\alpha},k'_{\alpha}}\left(n_{\uparrow}+ n_{\downarrow}\right) + K_{k_{\alpha},k'_{\alpha}}\right)c_{\vec{k'},\alpha}^{\dagger}c_{\vec{k},\alpha} + \left[\left(\frac{1}{2}W_{k'_{\alpha}}(n_{\uparrow}+n_{\downarrow})+Z_{k'_{\alpha}}\right) c_{\vec{-k'},\alpha}c_{\vec{k},\alpha}+ \mathrm{H.c.}\right]\right\rbrace
\end{split}
\end{equation}
with parameters 

\begin{equation}
\begin{split}
&K_{k_{\alpha},k'_{\alpha}} =  w \left(B(k_{\alpha})v^2(k_{\alpha}) + A(k_{\alpha}) |u(k_{\alpha})|^2 + B(k'_{\alpha})v^2(k'_{\alpha}) + A(k'_{\alpha}) |u(k'_{\alpha})|^2 \right) \\
& Z_{k_{\alpha}} = u(k_{\alpha})v(k_{\alpha})\left(A(k_{\alpha})-B(k_{\alpha})\right),
\end{split}
\end{equation}
and $H'_0$ shifts the energy level 

\begin{equation}
H'_0 = \sum_{\vec{k}}\left(K_{k_{x},k_{x}} \left(1-n_{\uparrow}-n_{\downarrow}\right) + \frac{1}{2}J_{k_{x},k_{x}} \left(n_{\uparrow}+n_{\downarrow}-4n_{\uparrow}n_{\downarrow}\right)\right).
\end{equation}

Of our interest is the Kondo exchange term \eqref{H_KAppen} projected to single-occupied states of impurity. In the limit of strong Hubbard interaction $U\gg|E(k_{\alpha})| $, the Kondo coupling $J_{k_{\alpha},k'_{\beta}}$ is approximated as 

\begin{equation}
\begin{split}
J_{k_{\alpha},k'_{\beta}} \simeq |w|^2 \left( \frac{2U}{\varepsilon_d(U+\varepsilon_d)}+\left(\varepsilon(k_{\alpha})+\varepsilon(k'_{\beta})\right)\frac{U(U+2\varepsilon_d)}{\varepsilon_d^2(U+\varepsilon_d)^2} \right).
\end{split}
\end{equation}

In the case of symmetric Anderson model, i.e., $\varepsilon_d = -U/2$, the Kondo exchange Hamiltonian becomes

\begin{equation}
H_K = \frac{J_K}{2}\sum_{\vec{k},\vec{k'}} \vec{S} \cdot \left(c^{\dagger}_{\vec{k'},\alpha} \boldsymbol{\sigma}_{\alpha\beta}c_{\vec{k},\beta}\right),
\end{equation}
where $J_K = \frac{8|w|^2}{U}$ and $\vec{S} =\frac{1}{2} d^{\dagger}_{\alpha} \boldsymbol{\tau}_{\alpha\beta}d_{\beta}$. The same exchange coupling is also obtained for a magnetic impurity in an $s$-wave superconductor \cite{Salomaa1988}. 

\end{widetext}

\newpage

\begin{thebibliography}{40}%
\makeatletter
\providecommand \@ifxundefined [1]{%
 \@ifx{#1\undefined}
}%
\providecommand \@ifnum [1]{%
 \ifnum #1\expandafter \@firstoftwo
 \else \expandafter \@secondoftwo
 \fi
}%
\providecommand \@ifx [1]{%
 \ifx #1\expandafter \@firstoftwo
 \else \expandafter \@secondoftwo
 \fi
}%
\providecommand \natexlab [1]{#1}%
\providecommand \enquote  [1]{``#1''}%
\providecommand \bibnamefont  [1]{#1}%
\providecommand \bibfnamefont [1]{#1}%
\providecommand \citenamefont [1]{#1}%
\providecommand \href@noop [0]{\@secondoftwo}%
\providecommand \href [0]{\begingroup \@sanitize@url \@href}%
\providecommand \@href[1]{\@@startlink{#1}\@@href}%
\providecommand \@@href[1]{\endgroup#1\@@endlink}%
\providecommand \@sanitize@url [0]{\catcode `\\12\catcode `\$12\catcode
  `\&12\catcode `\#12\catcode `\^12\catcode `\_12\catcode `\%12\relax}%
\providecommand \@@startlink[1]{}%
\providecommand \@@endlink[0]{}%
\providecommand \url  [0]{\begingroup\@sanitize@url \@url }%
\providecommand \@url [1]{\endgroup\@href {#1}{\urlprefix }}%
\providecommand \urlprefix  [0]{URL }%
\providecommand \Eprint [0]{\href }%
\providecommand \doibase [0]{https://doi.org/}%
\providecommand \selectlanguage [0]{\@gobble}%
\providecommand \bibinfo  [0]{\@secondoftwo}%
\providecommand \bibfield  [0]{\@secondoftwo}%
\providecommand \translation [1]{[#1]}%
\providecommand \BibitemOpen [0]{}%
\providecommand \bibitemStop [0]{}%
\providecommand \bibitemNoStop [0]{.\EOS\space}%
\providecommand \EOS [0]{\spacefactor3000\relax}%
\providecommand \BibitemShut  [1]{\csname bibitem#1\endcsname}%
\let\auto@bib@innerbib\@empty
\bibitem [{\citenamefont {Continentino}(1994)}]{Continentino1994}%
  \BibitemOpen
  \bibfield  {author} {\bibinfo {author} {\bibfnamefont {M.~A.}\ \bibnamefont
  {Continentino}},\ }\bibfield  {title} {\bibinfo {title} {Quantum scaling in
  many-body systems},\ }\href
  {https://doi.org/https://doi.org/10.1016/0370-1573(94)90112-0} {\bibfield
  {journal} {\bibinfo  {journal} {Physics Reports}\ }\textbf {\bibinfo {volume}
  {239}},\ \bibinfo {pages} {179} (\bibinfo {year} {1994})}\BibitemShut
  {NoStop}%
\bibitem [{\citenamefont {Sachdev}(2011)}]{sachdev:Book2011}%
  \BibitemOpen
  \bibfield  {author} {\bibinfo {author} {\bibfnamefont {S.}~\bibnamefont
  {Sachdev}},\ }\href {https://doi.org/10.1017/CBO9780511973765} {\emph
  {\bibinfo {title} {Quantum Phase Transitions}}},\ \bibinfo {edition} {2nd}\
  ed.\ (\bibinfo  {publisher} {Cambridge University Press},\ \bibinfo {year}
  {2011})\BibitemShut {NoStop}%
\bibitem [{\citenamefont {Hertz}(1976)}]{Hertz1976}%
  \BibitemOpen
  \bibfield  {author} {\bibinfo {author} {\bibfnamefont {J.~A.}\ \bibnamefont
  {Hertz}},\ }\bibfield  {title} {\bibinfo {title} {Quantum critical
  phenomena},\ }\href {https://doi.org/10.1103/PhysRevB.14.1165} {\bibfield
  {journal} {\bibinfo  {journal} {Phys. Rev. B}\ }\textbf {\bibinfo {volume}
  {14}},\ \bibinfo {pages} {1165} (\bibinfo {year} {1976})}\BibitemShut
  {NoStop}%
\bibitem [{\citenamefont {Millis}(1993)}]{Millis1993}%
  \BibitemOpen
  \bibfield  {author} {\bibinfo {author} {\bibfnamefont {A.~J.}\ \bibnamefont
  {Millis}},\ }\bibfield  {title} {\bibinfo {title} {Effect of a nonzero
  temperature on quantum critical points in itinerant fermion systems},\ }\href
  {https://doi.org/10.1103/PhysRevB.48.7183} {\bibfield  {journal} {\bibinfo
  {journal} {Phys. Rev. B}\ }\textbf {\bibinfo {volume} {48}},\ \bibinfo
  {pages} {7183} (\bibinfo {year} {1993})}\BibitemShut {NoStop}%
\bibitem [{\citenamefont {Si}\ \emph {et~al.}(2001)\citenamefont {Si},
  \citenamefont {Rabello}, \citenamefont {Ingersent},\ and\ \citenamefont
  {Smith}}]{Qimiao2001}%
  \BibitemOpen
  \bibfield  {author} {\bibinfo {author} {\bibfnamefont {Q.}~\bibnamefont
  {Si}}, \bibinfo {author} {\bibfnamefont {S.}~\bibnamefont {Rabello}},
  \bibinfo {author} {\bibfnamefont {K.}~\bibnamefont {Ingersent}},\ and\
  \bibinfo {author} {\bibfnamefont {J.~L.}\ \bibnamefont {Smith}},\ }\bibfield
  {title} {\bibinfo {title} {Locally critical quantum phase transitions in
  strongly correlated metals},\ }\href {https://doi.org/10.1038/35101507}
  {\bibfield  {journal} {\bibinfo  {journal} {Nature}\ }\textbf {\bibinfo
  {volume} {413}},\ \bibinfo {pages} {804} (\bibinfo {year}
  {2001})}\BibitemShut {NoStop}%
\bibitem [{\citenamefont {Coleman}\ and\ \citenamefont
  {Pepin}(2002)}]{Coleman2002}%
  \BibitemOpen
  \bibfield  {author} {\bibinfo {author} {\bibfnamefont {P.}~\bibnamefont
  {Coleman}}\ and\ \bibinfo {author} {\bibfnamefont {C.}~\bibnamefont
  {Pepin}},\ }\bibfield  {title} {\bibinfo {title} {What is the fate of the
  heavy electron at a quantum critical point?},\ }\href
  {https://doi.org/10.1016/S0921-4526(01)01342-4} {\bibfield  {journal}
  {\bibinfo  {journal} {Physica B: Condensed Matter}\ }\textbf {\bibinfo
  {volume} {312-313}},\ \bibinfo {pages} {383} (\bibinfo {year}
  {2002})}\BibitemShut {NoStop}%
\bibitem [{\citenamefont {Coleman}\ and\ \citenamefont
  {Schofield}(2005)}]{Coleman:Nature2005}%
  \BibitemOpen
  \bibfield  {author} {\bibinfo {author} {\bibfnamefont {P.}~\bibnamefont
  {Coleman}}\ and\ \bibinfo {author} {\bibfnamefont {A.~J.}\ \bibnamefont
  {Schofield}},\ }\bibfield  {title} {\bibinfo {title} {Quantum criticality},\
  }\href {https://doi.org/10.1038/nature03279} {\bibfield  {journal} {\bibinfo
  {journal} {Nature}\ }\textbf {\bibinfo {volume} {433}},\ \bibinfo {pages}
  {226} (\bibinfo {year} {2005})}\BibitemShut {NoStop}%
\bibitem [{\citenamefont {Ruderman}\ and\ \citenamefont
  {Kittel}(1954)}]{Ruderman1954}%
  \BibitemOpen
  \bibfield  {author} {\bibinfo {author} {\bibfnamefont {M.~A.}\ \bibnamefont
  {Ruderman}}\ and\ \bibinfo {author} {\bibfnamefont {C.}~\bibnamefont
  {Kittel}},\ }\bibfield  {title} {\bibinfo {title} {Indirect exchange coupling
  of nuclear magnetic moments by conduction electrons},\ }\href
  {https://doi.org/10.1103/PhysRev.96.99} {\bibfield  {journal} {\bibinfo
  {journal} {Phys. Rev.}\ }\textbf {\bibinfo {volume} {96}},\ \bibinfo {pages}
  {99} (\bibinfo {year} {1954})}\BibitemShut {NoStop}%
\bibitem [{\citenamefont {Kasuya}(1956)}]{Kasuya1956}%
  \BibitemOpen
  \bibfield  {author} {\bibinfo {author} {\bibfnamefont {T.}~\bibnamefont
  {Kasuya}},\ }\bibfield  {title} {\bibinfo {title} {{A Theory of Metallic
  Ferro- and Antiferromagnetism on Zener's Model}},\ }\href
  {https://doi.org/10.1143/PTP.16.45} {\bibfield  {journal} {\bibinfo
  {journal} {Progress of Theoretical Physics}\ }\textbf {\bibinfo {volume}
  {16}},\ \bibinfo {pages} {45} (\bibinfo {year} {1956})},\ \Eprint
  {https://arxiv.org/abs/https://academic.oup.com/ptp/article-pdf/16/1/45/5266722/16-1-45.pdf}
  {https://academic.oup.com/ptp/article-pdf/16/1/45/5266722/16-1-45.pdf}
  \BibitemShut {NoStop}%
\bibitem [{\citenamefont {Yosida}(1957)}]{Yosida1957}%
  \BibitemOpen
  \bibfield  {author} {\bibinfo {author} {\bibfnamefont {K.}~\bibnamefont
  {Yosida}},\ }\bibfield  {title} {\bibinfo {title} {Magnetic properties of
  cu-mn alloys},\ }\href {https://doi.org/10.1103/PhysRev.106.893} {\bibfield
  {journal} {\bibinfo  {journal} {Phys. Rev.}\ }\textbf {\bibinfo {volume}
  {106}},\ \bibinfo {pages} {893} (\bibinfo {year} {1957})}\BibitemShut
  {NoStop}%
\bibitem [{\citenamefont {Coleman}\ and\ \citenamefont
  {Andrei}(1989)}]{Coleman1989}%
  \BibitemOpen
  \bibfield  {author} {\bibinfo {author} {\bibfnamefont {P.}~\bibnamefont
  {Coleman}}\ and\ \bibinfo {author} {\bibfnamefont {N.}~\bibnamefont
  {Andrei}},\ }\bibfield  {title} {\bibinfo {title} {Kondo-stabilised spin
  liquids and heavy fermion superconductivity},\ }\href
  {https://doi.org/10.1088/0953-8984/1/26/003} {\bibfield  {journal} {\bibinfo
  {journal} {Journal of Physics: Condensed Matter}\ }\textbf {\bibinfo {volume}
  {1}},\ \bibinfo {pages} {4057} (\bibinfo {year} {1989})}\BibitemShut
  {NoStop}%
\bibitem [{\citenamefont {Hewson}(1993)}]{Hewson1993}%
  \BibitemOpen
  \bibfield  {author} {\bibinfo {author} {\bibfnamefont {A.}~\bibnamefont
  {Hewson}},\ }\bibfield  {title} {\bibinfo {title} {The kondo problem to heavy
  fermions},\ }\href@noop {} {\bibfield  {journal} {\bibinfo  {journal}
  {Cambridge University Press}\ } (\bibinfo {year} {1993})}\BibitemShut
  {NoStop}%
\bibitem [{\citenamefont {Coleman}(2007)}]{Coleman:Book2007}%
  \BibitemOpen
  \bibfield  {author} {\bibinfo {author} {\bibfnamefont {P.}~\bibnamefont
  {Coleman}},\ }\bibinfo {title} {Heavy fermions: Electrons at the edge of
  magnetism},\ in\ \href
  {https://doi.org/https://doi.org/10.1002/9780470022184.hmm105} {\emph
  {\bibinfo {booktitle} {Handbook of Magnetism and Advanced Magnetic
  Materials}}}\ (\bibinfo  {publisher} {John Wiley and Sons, Ltd},\ \bibinfo
  {year} {2007})\BibitemShut {NoStop}%
\bibitem [{\citenamefont {Coleman}(2015)}]{Coleman:Book2015}%
  \BibitemOpen
  \bibfield  {author} {\bibinfo {author} {\bibfnamefont {P.}~\bibnamefont
  {Coleman}},\ }\bibinfo {title} {Introduction},\ in\ \href
  {https://doi.org/10.1017/CBO9781139020916.002} {\emph {\bibinfo {booktitle}
  {Introduction to Many-Body Physics}}}\ (\bibinfo  {publisher} {Cambridge
  University Press},\ \bibinfo {year} {2015})\BibitemShut {NoStop}%
\bibitem [{\citenamefont {Luttinger}(1960)}]{Luttinger1960}%
  \BibitemOpen
  \bibfield  {author} {\bibinfo {author} {\bibfnamefont {J.~M.}\ \bibnamefont
  {Luttinger}},\ }\bibfield  {title} {\bibinfo {title} {Fermi surface and some
  simple equilibrium properties of a system of interacting fermions},\ }\href
  {https://doi.org/10.1103/PhysRev.119.1153} {\bibfield  {journal} {\bibinfo
  {journal} {Phys. Rev.}\ }\textbf {\bibinfo {volume} {119}},\ \bibinfo {pages}
  {1153} (\bibinfo {year} {1960})}\BibitemShut {NoStop}%
\bibitem [{\citenamefont {Senthil}\ \emph {et~al.}(2003)\citenamefont
  {Senthil}, \citenamefont {Sachdev},\ and\ \citenamefont
  {Vojta}}]{Senthil:PRL2003}%
  \BibitemOpen
  \bibfield  {author} {\bibinfo {author} {\bibfnamefont {T.}~\bibnamefont
  {Senthil}}, \bibinfo {author} {\bibfnamefont {S.}~\bibnamefont {Sachdev}},\
  and\ \bibinfo {author} {\bibfnamefont {M.}~\bibnamefont {Vojta}},\ }\bibfield
   {title} {\bibinfo {title} {Fractionalized fermi liquids},\ }\href
  {https://doi.org/10.1103/PhysRevLett.90.216403} {\bibfield  {journal}
  {\bibinfo  {journal} {Phys. Rev. Lett.}\ }\textbf {\bibinfo {volume} {90}},\
  \bibinfo {pages} {216403} (\bibinfo {year} {2003})}\BibitemShut {NoStop}%
\bibitem [{\citenamefont {Senthil}\ and\ \citenamefont
  {Fisher}(2000)}]{Senthil:prb1999}%
  \BibitemOpen
  \bibfield  {author} {\bibinfo {author} {\bibfnamefont {T.}~\bibnamefont
  {Senthil}}\ and\ \bibinfo {author} {\bibfnamefont {M.~P.~A.}\ \bibnamefont
  {Fisher}},\ }\bibfield  {title} {\bibinfo {title} {${Z}_{2}$ gauge theory of
  electron fractionalization in strongly correlated systems},\ }\href
  {https://doi.org/10.1103/PhysRevB.62.7850} {\bibfield  {journal} {\bibinfo
  {journal} {Phys. Rev. B}\ }\textbf {\bibinfo {volume} {62}},\ \bibinfo
  {pages} {7850} (\bibinfo {year} {2000})}\BibitemShut {NoStop}%
\bibitem [{\citenamefont {Yu}(1965)}]{Yu1965}%
  \BibitemOpen
  \bibfield  {author} {\bibinfo {author} {\bibfnamefont {L.}~\bibnamefont
  {Yu}},\ }\bibfield  {title} {\bibinfo {title} {Bound state in superconductors
  with paramagnetic impurities},\ }\href@noop {} {\bibfield  {journal}
  {\bibinfo  {journal} {Acta Phys. Sin}\ }\textbf {\bibinfo {volume} {21}},\
  \bibinfo {pages} {75} (\bibinfo {year} {1965})}\BibitemShut {NoStop}%
\bibitem [{\citenamefont {Shiba}(1968)}]{Shiba1968}%
  \BibitemOpen
  \bibfield  {author} {\bibinfo {author} {\bibfnamefont {H.}~\bibnamefont
  {Shiba}},\ }\bibfield  {title} {\bibinfo {title} {{Classical Spins in
  Superconductors}},\ }\href {https://doi.org/10.1143/PTP.40.435} {\bibfield
  {journal} {\bibinfo  {journal} {Progress of Theoretical Physics}\ }\textbf
  {\bibinfo {volume} {40}},\ \bibinfo {pages} {435} (\bibinfo {year} {1968})},\
  \Eprint
  {https://arxiv.org/abs/https://academic.oup.com/ptp/article-pdf/40/3/435/5185550/40-3-435.pdf}
  {https://academic.oup.com/ptp/article-pdf/40/3/435/5185550/40-3-435.pdf}
  \BibitemShut {NoStop}%
\bibitem [{\citenamefont {Rusinov}(1968)}]{Rusinov1968}%
  \BibitemOpen
  \bibfield  {author} {\bibinfo {author} {\bibfnamefont {A.}~\bibnamefont
  {Rusinov}},\ }\bibfield  {title} {\bibinfo {title} {Superconductivity near a
  paramagnetic impurity, pis’ ma zh},\ }\href@noop {} {\bibfield  {journal}
  {\bibinfo  {journal} {Eksp. Teor. Fiz}\ }\textbf {\bibinfo {volume} {9}},\
  \bibinfo {pages} {146} (\bibinfo {year} {1968})}\BibitemShut {NoStop}%
\bibitem [{\citenamefont {Nadj-Perge}\ \emph {et~al.}(2014)\citenamefont
  {Nadj-Perge}, \citenamefont {Drozdov}, \citenamefont {Li}, \citenamefont
  {Chen}, \citenamefont {Jeon}, \citenamefont {Seo}, \citenamefont {MacDonald},
  \citenamefont {Bernevig},\ and\ \citenamefont {Yazdani}}]{Yazdani2014}%
  \BibitemOpen
  \bibfield  {author} {\bibinfo {author} {\bibfnamefont {S.}~\bibnamefont
  {Nadj-Perge}}, \bibinfo {author} {\bibfnamefont {I.~K.}\ \bibnamefont
  {Drozdov}}, \bibinfo {author} {\bibfnamefont {J.}~\bibnamefont {Li}},
  \bibinfo {author} {\bibfnamefont {H.}~\bibnamefont {Chen}}, \bibinfo {author}
  {\bibfnamefont {S.}~\bibnamefont {Jeon}}, \bibinfo {author} {\bibfnamefont
  {J.}~\bibnamefont {Seo}}, \bibinfo {author} {\bibfnamefont {A.~H.}\
  \bibnamefont {MacDonald}}, \bibinfo {author} {\bibfnamefont {B.~A.}\
  \bibnamefont {Bernevig}},\ and\ \bibinfo {author} {\bibfnamefont
  {A.}~\bibnamefont {Yazdani}},\ }\bibfield  {title} {\bibinfo {title}
  {Observation of majorana fermions in ferromagnetic atomic chains on a
  superconductor},\ }\href {https://doi.org/10.1126/science.1259327} {\bibfield
   {journal} {\bibinfo  {journal} {Science}\ }\textbf {\bibinfo {volume}
  {346}},\ \bibinfo {pages} {602} (\bibinfo {year} {2014})},\ \Eprint
  {https://arxiv.org/abs/https://www.science.org/doi/pdf/10.1126/science.1259327}
  {https://www.science.org/doi/pdf/10.1126/science.1259327} \BibitemShut
  {NoStop}%
\bibitem [{\citenamefont {Kitaev}(2001)}]{Kitaev2001}%
  \BibitemOpen
  \bibfield  {author} {\bibinfo {author} {\bibfnamefont {A.~Y.}\ \bibnamefont
  {Kitaev}},\ }\bibfield  {title} {\bibinfo {title} {Unpaired majorana fermions
  in quantum wires},\ }\href@noop {} {\bibfield  {journal} {\bibinfo  {journal}
  {Physics-Uspekhi}\ }\textbf {\bibinfo {volume} {44}},\ \bibinfo {pages} {131}
  (\bibinfo {year} {2001})}\BibitemShut {NoStop}%
\bibitem [{\citenamefont {Pientka}\ \emph {et~al.}(2013)\citenamefont
  {Pientka}, \citenamefont {Glazman},\ and\ \citenamefont {von
  Oppen}}]{Pientka2013}%
  \BibitemOpen
  \bibfield  {author} {\bibinfo {author} {\bibfnamefont {F.}~\bibnamefont
  {Pientka}}, \bibinfo {author} {\bibfnamefont {L.~I.}\ \bibnamefont
  {Glazman}},\ and\ \bibinfo {author} {\bibfnamefont {F.}~\bibnamefont {von
  Oppen}},\ }\bibfield  {title} {\bibinfo {title} {Topological superconducting
  phase in helical shiba chains},\ }\href
  {https://doi.org/10.1103/PhysRevB.88.155420} {\bibfield  {journal} {\bibinfo
  {journal} {Phys. Rev. B}\ }\textbf {\bibinfo {volume} {88}},\ \bibinfo
  {pages} {155420} (\bibinfo {year} {2013})}\BibitemShut {NoStop}%
\bibitem [{\citenamefont {Hsieh}\ \emph {et~al.}(2017)\citenamefont {Hsieh},
  \citenamefont {Lu},\ and\ \citenamefont {Ludwig}}]{Hsieh2017}%
  \BibitemOpen
  \bibfield  {author} {\bibinfo {author} {\bibfnamefont {T.~H.}\ \bibnamefont
  {Hsieh}}, \bibinfo {author} {\bibfnamefont {Y.-M.}\ \bibnamefont {Lu}},\ and\
  \bibinfo {author} {\bibfnamefont {A.~W.~W.}\ \bibnamefont {Ludwig}},\
  }\bibfield  {title} {\bibinfo {title} {Topological bootstrap:
  Fractionalization from kondo coupling},\ }\bibfield  {journal} {\bibinfo
  {journal} {Science Advances}\ }\textbf {\bibinfo {volume} {3}},\ \href
  {https://doi.org/10.1126/sciadv.1700729} {10.1126/sciadv.1700729} (\bibinfo
  {year} {2017}),\ \Eprint
  {https://arxiv.org/abs/https://advances.sciencemag.org/content/3/10/e1700729.full.pdf}
  {https://advances.sciencemag.org/content/3/10/e1700729.full.pdf} \BibitemShut
  {NoStop}%
\bibitem [{\citenamefont {Mohammadi}\ and\ \citenamefont
  {Kargarian}(2022)}]{Mohammadi2022}%
  \BibitemOpen
  \bibfield  {author} {\bibinfo {author} {\bibfnamefont {F.}~\bibnamefont
  {Mohammadi}}\ and\ \bibinfo {author} {\bibfnamefont {M.}~\bibnamefont
  {Kargarian}},\ }\bibfield  {title} {\bibinfo {title} {Designing
  ${\mathbb{z}}_{2}$ and
  ${\mathbb{z}}_{2}\ifmmode\times\else\texttimes\fi{}{\mathbb{z}}_{2}$
  topological orders in networks of majorana bound states},\ }\href
  {https://doi.org/10.1103/PhysRevB.105.165107} {\bibfield  {journal} {\bibinfo
   {journal} {Phys. Rev. B}\ }\textbf {\bibinfo {volume} {105}},\ \bibinfo
  {pages} {165107} (\bibinfo {year} {2022})}\BibitemShut {NoStop}%
\bibitem [{\citenamefont {Hsieh}\ \emph {et~al.}(2016)\citenamefont {Hsieh},
  \citenamefont {Ishizuka}, \citenamefont {Balents},\ and\ \citenamefont
  {Hughes}}]{Hsieh2016}%
  \BibitemOpen
  \bibfield  {author} {\bibinfo {author} {\bibfnamefont {T.~H.}\ \bibnamefont
  {Hsieh}}, \bibinfo {author} {\bibfnamefont {H.}~\bibnamefont {Ishizuka}},
  \bibinfo {author} {\bibfnamefont {L.}~\bibnamefont {Balents}},\ and\ \bibinfo
  {author} {\bibfnamefont {T.~L.}\ \bibnamefont {Hughes}},\ }\bibfield  {title}
  {\bibinfo {title} {Bulk topological proximity effect},\ }\href
  {https://doi.org/10.1103/PhysRevLett.116.086802} {\bibfield  {journal}
  {\bibinfo  {journal} {Phys. Rev. Lett.}\ }\textbf {\bibinfo {volume} {116}},\
  \bibinfo {pages} {086802} (\bibinfo {year} {2016})}\BibitemShut {NoStop}%
\bibitem [{\citenamefont {Coleman}\ \emph {et~al.}(2005)\citenamefont
  {Coleman}, \citenamefont {Marston},\ and\ \citenamefont
  {Schofield}}]{Coleman:prb2005}%
  \BibitemOpen
  \bibfield  {author} {\bibinfo {author} {\bibfnamefont {P.}~\bibnamefont
  {Coleman}}, \bibinfo {author} {\bibfnamefont {J.~B.}\ \bibnamefont
  {Marston}},\ and\ \bibinfo {author} {\bibfnamefont {A.~J.}\ \bibnamefont
  {Schofield}},\ }\bibfield  {title} {\bibinfo {title} {Transport anomalies in
  a simplified model for a heavy-electron quantum critical point},\ }\href
  {https://doi.org/10.1103/PhysRevB.72.245111} {\bibfield  {journal} {\bibinfo
  {journal} {Phys. Rev. B}\ }\textbf {\bibinfo {volume} {72}},\ \bibinfo
  {pages} {245111} (\bibinfo {year} {2005})}\BibitemShut {NoStop}%
\bibitem [{\citenamefont {Salomaa}(1988)}]{Salomaa1988}%
  \BibitemOpen
  \bibfield  {author} {\bibinfo {author} {\bibfnamefont {M.~M.}\ \bibnamefont
  {Salomaa}},\ }\bibfield  {title} {\bibinfo {title} {Schrieffer-wolff
  transformation for the anderson hamiltonian in a superconductor},\ }\href
  {https://doi.org/10.1103/PhysRevB.37.9312} {\bibfield  {journal} {\bibinfo
  {journal} {Phys. Rev. B}\ }\textbf {\bibinfo {volume} {37}},\ \bibinfo
  {pages} {9312} (\bibinfo {year} {1988})}\BibitemShut {NoStop}%
\bibitem [{\citenamefont {Kotliar}\ and\ \citenamefont
  {Liu}(1988)}]{Kotliar:prb1988}%
  \BibitemOpen
  \bibfield  {author} {\bibinfo {author} {\bibfnamefont {G.}~\bibnamefont
  {Kotliar}}\ and\ \bibinfo {author} {\bibfnamefont {J.}~\bibnamefont {Liu}},\
  }\bibfield  {title} {\bibinfo {title} {Superexchange mechanism and d-wave
  superconductivity},\ }\href {https://doi.org/10.1103/PhysRevB.38.5142}
  {\bibfield  {journal} {\bibinfo  {journal} {Phys. Rev. B}\ }\textbf {\bibinfo
  {volume} {38}},\ \bibinfo {pages} {5142} (\bibinfo {year}
  {1988})}\BibitemShut {NoStop}%
\bibitem [{\citenamefont {Wen}(2003)}]{Wen:PRL2003}%
  \BibitemOpen
  \bibfield  {author} {\bibinfo {author} {\bibfnamefont {X.-G.}\ \bibnamefont
  {Wen}},\ }\bibfield  {title} {\bibinfo {title} {Quantum orders in an exact
  soluble model},\ }\href {https://doi.org/10.1103/PhysRevLett.90.016803}
  {\bibfield  {journal} {\bibinfo  {journal} {Phys. Rev. Lett.}\ }\textbf
  {\bibinfo {volume} {90}},\ \bibinfo {pages} {016803} (\bibinfo {year}
  {2003})}\BibitemShut {NoStop}%
\bibitem [{\citenamefont {Wen}(2002)}]{Wen:PRB2003}%
  \BibitemOpen
  \bibfield  {author} {\bibinfo {author} {\bibfnamefont {X.-G.}\ \bibnamefont
  {Wen}},\ }\bibfield  {title} {\bibinfo {title} {Quantum orders and symmetric
  spin liquids},\ }\href {https://doi.org/10.1103/PhysRevB.65.165113}
  {\bibfield  {journal} {\bibinfo  {journal} {Phys. Rev. B}\ }\textbf {\bibinfo
  {volume} {65}},\ \bibinfo {pages} {165113} (\bibinfo {year}
  {2002})}\BibitemShut {NoStop}%
\bibitem [{\citenamefont {Moca}\ \emph {et~al.}(2021)\citenamefont {Moca},
  \citenamefont {Weymann}, \citenamefont {Werner},\ and\ \citenamefont
  {Zar\'and}}]{Moca:PRL2021}%
  \BibitemOpen
  \bibfield  {author} {\bibinfo {author} {\bibfnamefont {C.~u. u. u. u. P.
  m.~c.}\ \bibnamefont {Moca}}, \bibinfo {author} {\bibfnamefont
  {I.}~\bibnamefont {Weymann}}, \bibinfo {author} {\bibfnamefont {M.~A.}\
  \bibnamefont {Werner}},\ and\ \bibinfo {author} {\bibfnamefont
  {G.}~\bibnamefont {Zar\'and}},\ }\bibfield  {title} {\bibinfo {title} {Kondo
  cloud in a superconductor},\ }\href
  {https://doi.org/10.1103/PhysRevLett.127.186804} {\bibfield  {journal}
  {\bibinfo  {journal} {Phys. Rev. Lett.}\ }\textbf {\bibinfo {volume} {127}},\
  \bibinfo {pages} {186804} (\bibinfo {year} {2021})}\BibitemShut {NoStop}%
\bibitem [{\citenamefont {Yarloo}\ \emph {et~al.}(2018)\citenamefont {Yarloo},
  \citenamefont {Langari},\ and\ \citenamefont {Vaezi}}]{Yarloo2018}%
  \BibitemOpen
  \bibfield  {author} {\bibinfo {author} {\bibfnamefont {H.}~\bibnamefont
  {Yarloo}}, \bibinfo {author} {\bibfnamefont {A.}~\bibnamefont {Langari}},\
  and\ \bibinfo {author} {\bibfnamefont {A.}~\bibnamefont {Vaezi}},\ }\bibfield
   {title} {\bibinfo {title} {Anyonic self-induced disorder in a stabilizer
  code: Quasi many-body localization in a translational invariant model},\
  }\href {https://doi.org/10.1103/PhysRevB.97.054304} {\bibfield  {journal}
  {\bibinfo  {journal} {Phys. Rev. B}\ }\textbf {\bibinfo {volume} {97}},\
  \bibinfo {pages} {054304} (\bibinfo {year} {2018})}\BibitemShut {NoStop}%
\bibitem [{\citenamefont {Cozzini}\ \emph {et~al.}(2007)\citenamefont
  {Cozzini}, \citenamefont {Ionicioiu},\ and\ \citenamefont
  {Zanardi}}]{Cozzini2007}%
  \BibitemOpen
  \bibfield  {author} {\bibinfo {author} {\bibfnamefont {M.}~\bibnamefont
  {Cozzini}}, \bibinfo {author} {\bibfnamefont {R.}~\bibnamefont {Ionicioiu}},\
  and\ \bibinfo {author} {\bibfnamefont {P.}~\bibnamefont {Zanardi}},\
  }\bibfield  {title} {\bibinfo {title} {Quantum fidelity and quantum phase
  transitions in matrix product states},\ }\href
  {https://doi.org/10.1103/PhysRevB.76.104420} {\bibfield  {journal} {\bibinfo
  {journal} {Phys. Rev. B}\ }\textbf {\bibinfo {volume} {76}},\ \bibinfo
  {pages} {104420} (\bibinfo {year} {2007})}\BibitemShut {NoStop}%
\bibitem [{\citenamefont {Zhou}\ and\ \citenamefont
  {Barjaktarevic}(2007)}]{Zhou2007}%
  \BibitemOpen
  \bibfield  {author} {\bibinfo {author} {\bibfnamefont {H.}~\bibnamefont
  {Zhou}}\ and\ \bibinfo {author} {\bibfnamefont {J.~P.}\ \bibnamefont
  {Barjaktarevic}},\ }\bibfield  {title} {\bibinfo {title} {Fidelity and
  quantum phase transitions},\ }\href@noop {} {\bibfield  {journal} {\bibinfo
  {journal} {Journal of Physics A}\ }\textbf {\bibinfo {volume} {41}},\
  \bibinfo {pages} {412001} (\bibinfo {year} {2007})}\BibitemShut {NoStop}%
\bibitem [{\citenamefont {Zhou}\ \emph {et~al.}(2008)\citenamefont {Zhou},
  \citenamefont {Zhao},\ and\ \citenamefont {Li}}]{Zhou2008}%
  \BibitemOpen
  \bibfield  {author} {\bibinfo {author} {\bibfnamefont {H.-Q.}\ \bibnamefont
  {Zhou}}, \bibinfo {author} {\bibfnamefont {J.-H.}\ \bibnamefont {Zhao}},\
  and\ \bibinfo {author} {\bibfnamefont {B.}~\bibnamefont {Li}},\ }\bibfield
  {title} {\bibinfo {title} {Fidelity approach to quantum phase transitions:
  finite-size scaling for the quantum ising model in a transverse field},\
  }\href {https://doi.org/10.1088/1751-8113/41/49/492002} {\bibfield  {journal}
  {\bibinfo  {journal} {Journal of Physics A: Mathematical and Theoretical}\
  }\textbf {\bibinfo {volume} {41}},\ \bibinfo {pages} {492002} (\bibinfo
  {year} {2008})}\BibitemShut {NoStop}%
\bibitem [{\citenamefont {Holzhey}\ \emph {et~al.}(1994)\citenamefont
  {Holzhey}, \citenamefont {Larsen},\ and\ \citenamefont
  {Wilczek}}]{Holzhey1994}%
  \BibitemOpen
  \bibfield  {author} {\bibinfo {author} {\bibfnamefont {C.}~\bibnamefont
  {Holzhey}}, \bibinfo {author} {\bibfnamefont {F.}~\bibnamefont {Larsen}},\
  and\ \bibinfo {author} {\bibfnamefont {F.}~\bibnamefont {Wilczek}},\
  }\bibfield  {title} {\bibinfo {title} {Geometric and renormalized entropy in
  conformal field theory},\ }\href
  {https://doi.org/10.1016/0550-3213(94)90402-2} {\bibfield  {journal}
  {\bibinfo  {journal} {Nuclear Physics B}\ }\textbf {\bibinfo {volume}
  {424}},\ \bibinfo {pages} {443} (\bibinfo {year} {1994})}\BibitemShut
  {NoStop}%
\bibitem [{\citenamefont {Vidal}\ \emph {et~al.}(2003)\citenamefont {Vidal},
  \citenamefont {Latorre}, \citenamefont {Rico},\ and\ \citenamefont
  {Kitaev}}]{Vidal2003}%
  \BibitemOpen
  \bibfield  {author} {\bibinfo {author} {\bibfnamefont {G.}~\bibnamefont
  {Vidal}}, \bibinfo {author} {\bibfnamefont {J.~I.}\ \bibnamefont {Latorre}},
  \bibinfo {author} {\bibfnamefont {E.}~\bibnamefont {Rico}},\ and\ \bibinfo
  {author} {\bibfnamefont {A.}~\bibnamefont {Kitaev}},\ }\bibfield  {title}
  {\bibinfo {title} {Entanglement in quantum critical phenomena},\ }\href
  {https://doi.org/10.1103/PhysRevLett.90.227902} {\bibfield  {journal}
  {\bibinfo  {journal} {Phys. Rev. Lett.}\ }\textbf {\bibinfo {volume} {90}},\
  \bibinfo {pages} {227902} (\bibinfo {year} {2003})}\BibitemShut {NoStop}%
\bibitem [{\citenamefont {Calabrese}\ and\ \citenamefont
  {Cardy}(2009)}]{Calabrese2009}%
  \BibitemOpen
  \bibfield  {author} {\bibinfo {author} {\bibfnamefont {P.}~\bibnamefont
  {Calabrese}}\ and\ \bibinfo {author} {\bibfnamefont {J.}~\bibnamefont
  {Cardy}},\ }\bibfield  {title} {\bibinfo {title} {Entanglement entropy and
  conformal field theory},\ }\href
  {https://doi.org/10.1088/1751-8113/42/50/504005} {\bibfield  {journal}
  {\bibinfo  {journal} {Journal of Physics A: Mathematical and Theoretical}\
  }\textbf {\bibinfo {volume} {42}},\ \bibinfo {pages} {504005} (\bibinfo
  {year} {2009})}\BibitemShut {NoStop}%
\bibitem [{\citenamefont {Verresen}\ \emph {et~al.}(2017)\citenamefont
  {Verresen}, \citenamefont {Moessner},\ and\ \citenamefont
  {Pollmann}}]{Verresen2017}%
  \BibitemOpen
  \bibfield  {author} {\bibinfo {author} {\bibfnamefont {R.}~\bibnamefont
  {Verresen}}, \bibinfo {author} {\bibfnamefont {R.}~\bibnamefont {Moessner}},\
  and\ \bibinfo {author} {\bibfnamefont {F.}~\bibnamefont {Pollmann}},\
  }\bibfield  {title} {\bibinfo {title} {One-dimensional symmetry protected
  topological phases and their transitions},\ }\href
  {https://doi.org/10.1103/PhysRevB.96.165124} {\bibfield  {journal} {\bibinfo
  {journal} {Phys. Rev. B}\ }\textbf {\bibinfo {volume} {96}},\ \bibinfo
  {pages} {165124} (\bibinfo {year} {2017})}\BibitemShut {NoStop}%
\end{thebibliography}
%

\end{document}